\documentclass[preprint]{revtex4}

\usepackage{amsfonts}
\usepackage{amsmath}
\usepackage{amssymb}
\usepackage{graphicx}
\usepackage{color}

\begin{document}

\title{Does stability of relativistic dissipative
fluid dynamics imply causality?}
\author{Shi Pu$^{a,c}$}
\author{Tomoi Koide$^{b}$}
\author{Dirk H.\ Rischke$^{a,b}$}

\affiliation{$^{a}$Institut f\"ur Theoretische Physik,
Johann Wolfgang Goethe-Universit\"at,
Max-von-Laue-Str.\ 1, D-60438 Frankfurt am Main, Germany}
\affiliation{$^{b}$Frankfurt Institute for Advanced Studies,
Ruth-Moufang-Str.\ 1, D-60438 Frankfurt am Main, Germany}
\affiliation{$^{c}$Department of Modern Physics, University of
Science and Technology of China, Hefei 230026, P.R.\ China}

\begin{abstract}
We investigate the causality and stability of
relativistic dissipative fluid dynamics in the absence of
conserved charges. We perform a linear stability analysis
in the rest frame of the fluid and find that the equations of
relativistic dissipative fluid dynamics are always stable.
We then perform
a linear stability analysis in a Lorentz-boosted frame.
Provided that the ratio of the
relaxation time for the shear stress tensor, $\tau_\pi$,
to the sound attenuation length, $\Gamma_s = 4\eta/3(\varepsilon+P)$,
fulfills a certain asymptotic
causality condition, the equations of motion give rise to stable solutions.
Although the group velocity associated with
perturbations may exceed the velocity of light
in a certain finite range of wavenumbers, we demonstrate
that this does not violate causality, as long as the asymptotic
causality condition is fulfilled.
Finally, we compute the characteristic velocities
and show that they remain below the velocity of light if
the ratio $\tau_\pi/\Gamma_s$
fulfills the asymptotic causality condition.
\end{abstract}

\maketitle

\section{Introduction}

Data from the Relativistic Heavy-Ion Collider (RHIC) on the
collective flow of matter in nucleus-nucleus collisions have
delivered a surprising result: the elliptic flow coefficient
$v_2$ is sufficiently large
\cite{Adams:2003zg,Adams:2003am,Sorensen:2003kp,Adler:2002pu}
to be compatible with calculations performed in the
framework of {\em ideal\/} fluid dynamics
\cite{review}.
This has given rise to the notion that ``RHIC physicists serve up
the perfect liquid'' \cite{press,Gyulassy:2004zy,Shuryak:2004cy}.

Of course, no real liquid can have zero viscosity:
for all weakly coupled theories, i.e., theories with well-defined
quasi-particles, in the dilute limit
there is a lower bound which one can derive
from the uncertainty principle
\cite{Danielewicz:1984ww}: the ratio of shear viscosity to
entropy density $\eta/s \agt 1/12$. For certain strongly
coupled theories without quasiparticles, there is
also a lower bound which can be obtained from the
AdS/CFT conjecture \cite{Kovtun:2004de},
$\eta/s \geq 1/(4 \pi)$, i.e., surprisingly
close to the bound for dilute, weakly coupled systems.

In order to see whether the shear viscosity of the hot and dense
matter created in nuclear collisions at RHIC is close to the lower
bound, one has to perform calculations in the framework of
relativistic {\em dissipative\/} fluid dynamics. This program has
only been recently initiated, but has already led to an
enormous activity in the literature
\cite{muronga,roma,luzum1,luzum2,Song:2007ux,Song:2007fn,chau,du,pasi,pratt,bha,mol,Betz:2008me,gue,dkkm1,dkkm2,dkkm3,dkkm4,dkkm5,knk}.

Fluid dynamics is an effective theory for the long-wavelength,
small-frequency modes of a given theory. In order to see this,
let us introduce three length scales:
(a) a microscopic length scale, $\ell_{\rm micro}$. In all
theories, at sufficiently large temperatures
this length scale can be defined as the
thermal wavelength $\lambda_{\rm th} \sim 1/T$.
In weakly coupled theories with well-defined quasi-particles,
this can be interpreted as the interparticle distance.
(b) A mesoscopic length scale, $\ell_{\rm meso}$.
In weakly coupled theories and in the dilute limit, this
can be identified with the mean-free path of particles
between collisions. In strongly coupled theories,
such a scale is not known and should be identified with
$\ell_{\rm micro}$.
(c) A macroscopic length scale,
$\ell_{\rm macro}$. This is
the scale over which the conserved densities
(e.g.\ the charge density, $n$, or the energy density,
$\varepsilon$) of the theory vary. Thus, $\ell_{\rm macro}^{-1}
\sim |\partial \varepsilon|/\varepsilon$, i.e.,
$\ell_{\rm macro}^{-1}$ is proportional to the
gradients of the conserved quantities.

We now define the quantity
$K \equiv \ell_{\rm meso}/\ell_{\rm macro}$.
For dilute systems, this quantity is identical to
the so-called Knudsen number.
If $K$ is sufficiently small,
fluid dynamics as an effective theory
can be derived in a controlled way as a
power series in terms $K$. Since $K \sim \ell_{\rm macro}^{-1}$,
this series expansion is equivalent to a gradient expansion.

To zeroth order in $K$, one obtains the equations of
ideal fluid dynamics.
To first order in $K$, one obtains the Navier-Stokes (NS)
equations. So-called second-order theories
contain terms of second order in $K$. Examples for the latter are the
Burnett equations \cite{samojeden}, the
Israel-Stewart equations for relativistic dissipative
fluid dynamics \cite{is}, the memory function theory
\cite{dkkm1,dkkm4}, extended thermodynamics
\cite{jou,dkkm4}, and others \cite{else}.
The main difference between first and second-order theories
is the velocity of signal propagation. The relativistic
NS equations allow for infinite signal propagation speeds
and are therefore acausal.
On the other hand, all second-order theories are considered
to be causal in the sense that all signal velocities are
smaller than the speed of light, provided that the
parameters of the theory are suitably chosen.

The stability and causality of fluid-dynamical theories are
usually studied around a hydrostatic state (i.e., for vanishing
macroscopic flow velocity) which is in thermodynamical
equilibrium. However, if a theory is stable around a
hydrostatic state, it does not necessarily imply that
it is stable in a state of nonzero flow velocity.
Following this idea, the stability and causality of first and
second-order fluid dynamics for a state with nonzero
background flow velocity (mathematically realized by
a Lorentz boost) were studied
for the case of nonzero bulk viscosity, but for
vanishing shear stress and heat flow in Ref.\ \cite{dkkm3}.
There it was found that causality and stability are
intimately related: for all parameters
considered, the theory becomes unstable
if and only if there is a mode which propagates faster than
the speed of light.

In this paper, we extend this analysis to the case
of nonvanishing shear viscosity in second-order theories
of relativistic dissipative fluid dynamics.
A similar analysis for a hydrostatic background
has already been done by Hiscock,
Lindblom, and Olson \cite{his,his2}, but they
discussed exclusively the low- and high-wavenumber
limits \cite{his2}.
As we shall show in this paper, their analysis
missed a divergence of the group velocity of a
shear mode at intermediate wavenumbers.
This anomalous behavior is generic, i.e., it
cannot be removed by tuning the parameters of the theory, e.g.,
the relaxation time for the shear stress tensor, $\tau_\pi$,
and the shear viscosity, $\eta$.
However, if the ratio $\tau_\pi/\Gamma_s$,
where $\Gamma_s = 2(D-2)\eta/[(D-1)(\varepsilon + P)]$
is the sound attenuation length in $D$ space-time dimensions,
is chosen such that
the large-momentum limit of the group velocity
associated with the perturbation remains
below the velocity of light (the so-called
asymptotic causality condition), one can
ensure that the divergence is restricted to
a finite range of momenta. It will be demonstrated that in this case,
the causality of the theory is not compromised.
On the other hand,
second-order fluid dynamics is always stable in the rest frame of the fluid,
even if we use a parameter set which violates the
asymptotic causality condition.

We also study the causality and stability for a state
with nonzero background flow velocity, i.e., in
a Lorentz-boosted frame. We find that the divergence
of the group velocity is removed. However,
depending on the boost velocity the group velocity of either
the shear or the sound mode may still exceed the
speed of light in a certain range of wavenumbers.
Nevertheless, provided that the ratio $\tau_\pi/\Gamma_s$
fulfills the asymptotic causality condition,
we can show that the equations are stable. In contrast
to the analysis in the rest frame, however, they become
unstable if the asymptotic causality condition is violated.
We shall demonstrate that if
the asymptotic causality condition is fulfilled, the causality
of the theory as a whole is not compromised. In this sense,
causality and stability are intimately related.

So far, the discussion was limited to
the fluid-dynamical equations
in the linear approximation. Therefore, we expect
the results to be valid for all versions
of second-order theories presently discussed in the
literature, since they differ only by nonlinear terms.
We also compute the characteristic velocities
for the so-called simplified IS equations \cite{Song:2007ux}
without linearizing these equations. Our analysis strongly
indicates that the characteristic velocities
remain below the velocity of light if
the ratio $\tau_\pi/\Gamma_s$ is chosen such that the
asymptotic causality condition is fulfilled.

The asymptotic causality condition implies that, for a given
$\Gamma_s \sim \eta$, $\tau_\pi$
must not be arbitrarily small. This explains why
relativistic NS theory is acausal, because there $\tau_\pi
\rightarrow 0$, while $\eta$ is non-zero.
It also implies that second-order
theories are not {\em per se\/} causal; they can violate
causality (and become unstable) if a too small value for
$\tau_\pi$ is chosen. The statement that second-order
theories automatically cure the shortcomings of NS theory is
therefore not true.

This paper is organized as follows. In Sec.\ \ref{II},
we discuss the causality and stability of the
linearized second-order
fluid-dynamical equations in the local rest frame.
We also extend this analysis to nonzero bulk viscosity
and show that the divergence of the group velocity still exists
in this case.
In Sec.\ \ref{III}, this discussion is
generalized to a Lorentz-boosted frame.
We discuss Lorentz boosts both
in and orthogonal to the direction of propagation
of the perturbation.
It will be demonstrated that superluminal group velocities
will not compromise the causality of the theory as long
as the asymptotic causality condition is fulfilled.
In Sec.\ \ref{IV},
we compute the characteristic velocities in the nonlinear
case. A summary of our results concludes this work
in Sec.\ \ref{V}. An
Appendix  contains details of our
calculations in Sec.\ \ref{IV}. The metric tensor is
$g^{\mu \nu} = {\rm diag} (+,-,-,-)$; our units are
$\hbar = c = k_B =1$.

\section{Stability in the rest frame} \label{II}

As mentioned in the Introduction,
there are several approaches to formulate a
second-order theory of relativistic dissipative fluids
\cite{is,dkkm1,dkkm3,dkkm4,jou,else}.
These approaches differ only by nonlinear (second-order)
terms. However, since we shall apply a linear
stability analysis in the following, these differences
vanish and all approaches lead to the same set of
linearized fluid-dynamical equations. In this work,
we do not consider any conserved charges and thus are
left with energy-momentum conservation,
\begin{equation}
\partial_\mu T^{\mu \nu} = 0\;,
\end{equation}
where
\begin{equation}
T^{\mu\nu} = \varepsilon \, u^{\mu}u^{\nu}
-(P+\Pi)\Delta^{\mu\nu}+\pi^{\mu\nu}
\end{equation}
is the energy-momentum tensor. Here,
$\varepsilon$ and $P$ are the energy density
and the pressure, while
$u^{\mu}$, $\Pi$, and $\pi^{\mu\nu}$ are the
fluid velocity, the bulk viscous pressure, and
the shear stress tensor, respectively.
We also introduced the projection operator
\begin{equation}
\Delta^{\mu\nu} = g^{\mu\nu} - u^{\mu}u^{\nu}\;,
\end{equation}
which projects onto the $(D-1)$-dimensional subspace
orthogonal to the fluid velocity.
We compute in the Landau frame \cite{LL},
where there is no energy flow in the local rest frame.

In second-order theories of relativistic dissipative
fluid dynamics, the bulk viscous pressure and the
shear stress tensor are determined from evolution equations.
In $D$ space-time dimensions ($D \ge 3$),
these equations are given by
\begin{subequations} \label{eqofmotion}
\begin{eqnarray}
\tau_{\Pi}\,\frac{d}{d\tau}\Pi+\Pi
& = & -\zeta\, \partial_{\mu}u^{\mu}\;,\label{eq:bulk case}\\
\tau_{\pi}\, P^{\mu\nu\alpha\beta}\,
\frac{d}{d\tau}\pi_{\alpha\beta}+\pi^{\mu\nu}
& = & 2 \eta \, P^{\mu\nu\alpha\beta}\,
\partial_{\alpha}u_{\beta}\;;  \label{eq:shear case}
\end{eqnarray}
\end{subequations}
possible other second-order terms \cite{Betz:2008me}
can be neglected for the
purpose of a linear stability analysis.
In Eqs.\ (\ref{eqofmotion}),
the comoving derivative is denoted by
$u^\mu \partial_\mu \equiv d/d\tau$.
The relaxation times for the bulk viscous pressure
and the shear stress tensor are denoted by
$\tau_\Pi$ and $\tau_\pi$, respectively.
The coefficients $\zeta,\, \eta$ are the bulk and shear
viscosities, respectively.
We also introduced the symmetric rank-four
projection operator
\begin{equation}
P^{\mu\nu\alpha\beta}
= \frac{1}{2}\left(\Delta^{\mu\alpha}\Delta^{\nu\beta}
+\Delta^{\nu\alpha}\Delta^{\mu\beta}\right)
-\frac{1}{D-1}\,\Delta^{\mu\nu}\Delta^{\alpha\beta}\; .
\end{equation}
The shear stress tensor is traceless $\pi^{\mu}{}_{\mu} = 0$
and orthogonal to the fluid velocity $u_{\mu}\pi^{\mu\nu} =0$.

The stability and causality of a relativistic
dissipative fluid with bulk viscous pressure only
have been investigated in Ref.\ \cite{dkkm3}.
Thus, for the sake of simplicity, we shall first
ignore the effects from bulk viscous pressure and discuss
the properties of the fluid-dynamical equations of motion
including only shear viscosity.
The interplay between shear and bulk viscosity
will be discussed afterwards.

\subsection{Shear viscosity only} \label{IIA}

For convenience, we introduce the following parameterization:
\begin{subequations} \label{param}
\begin{eqnarray}
\eta &=& a s\;, \\
\tau_{\pi} &=& \frac{\eta}{\varepsilon + P}\, b=\frac{ab}{T}\;,
\label{param_b}
\end{eqnarray}
\end{subequations}
where $s$ and $T$ are the entropy density and the temperature,
respectively. From the
second equation we obtain $\tau_\pi (\varepsilon + P)/\eta = b$.
The parametrization (\ref{param}) is motivated by the
leading-order
results for the causal shear viscosity coefficient and the
relaxation time obtained in Ref.\ \cite{knk} where the relation
$\tau_\pi = \eta/P$ was found. For a massless ideal gas equation of
state, $\varepsilon = (D-1)P$, this result is
reproduced by choosing $b=D$.

In this section, we discuss the stability of
second-order relativistic fluid dynamics in the local rest frame.
Following Ref.\ \cite{his,dkkm3}, let us introduce a
perturbation $\sim e^{i\omega t - i kx}$
around the hydrostatic equilibrium state,
\begin{subequations}
\begin{eqnarray}
\varepsilon &=& \varepsilon_0 + \delta\varepsilon
\, e^{i\omega t - ikx}\;, \\
\pi^{\mu\nu} &=& \pi^{\mu\nu}_0 + \delta \pi^{\mu\nu}
\, e^{i\omega t - ikx}\;, \\
u^{\mu} &=& u^{\mu}_0 + \delta u^{\mu}\, e^{i\omega t - ikx}\;,
\end{eqnarray}
\end{subequations}
where $\varepsilon_0 = {\rm const.}$,
$\pi^{\mu\nu}_0 = 0$, and $u^{\mu}_0 = (1,0,0,\ldots)$,
respectively.
In the linear approximation,
the velocity perturbation has no zeroth component,
\begin{equation}
\delta u^{\mu} = (0,\delta u^{1},\delta u^{2},
\ldots,\delta u^{D-1})\;,
\end{equation}
because $u^{\mu}u_{\mu} =1$.
Moreover, in the local rest frame, $\delta \pi^{0\nu} \equiv 0$
on account of the orthogonality condition
$u_\mu\pi^{\mu \nu} = 0$. Since $\pi^{\mu \nu}$ is traceless,
$\delta \pi^{(D-1)(D-1)}$ is not an independent variable.
Taking all of this into account,
the linearized fluid-dynamical equations can be written as
\begin{equation} \label{lineq}
AX=0\;,
\end{equation}
where
\begin{eqnarray*}
X & = & (\delta\varepsilon,\delta u^{1},\delta \pi^{11},
\delta u^{2},\delta \pi^{12},
\ldots,\delta u^{D-1},\delta \pi^{1(D-1)},\\
&   & \;\; \delta \pi^{22},\delta \pi^{33},\ldots,
\delta \pi^{(D-2)(D-2)},
\delta \pi^{23}, \delta \pi^{24},\ldots,
\delta \pi^{2(D-1)}, \delta \pi^{34},\ldots,
\delta \pi^{(D-2)(D-1)})^T\;.
\end{eqnarray*}
The matrix $A$ is expressed as
\begin{equation}
A=\left(
\begin{array}{cccc}
T&0&0&0\\
 0& B&0&0\\
 G& 0 & C&0\\
0 & 0 & 0 & E
\end{array}
\right)\;,
\end{equation}
with
\begin{subequations}
\begin{eqnarray}
T &=&
\left(\begin{array}{ccc}
i\omega & f_{1} & 0\\
-ikc_{s}^{2} & f_{2} & -ik\\
0 & \Gamma & f
\end{array}
\right)\;, \\
B &=& {\rm diag}(B_0, \ldots, B_0)_{(D-2)\times (D-2)}\;,
\;\; B_0 =
\left(\begin{array}{cc}
f_{2} & -ik\\
\Gamma_{1} & f
\end{array}
\right)
\;, \\
G &=& \left(
 \begin{array}{ccc}
0 & \Gamma_2 & 0 \\
 & \ldots &\\
 0 & \Gamma_2 & 0 \end{array}\right)_{(D-3)\times 3} \;, \\
C & = & {\rm diag}(f,\ldots, f)_{(D-3)\times(D-3)}\; , \\
E & = & {\rm diag}(f,\ldots,f)_{\frac{1}{2}(D-2)(D-3)
\times\frac{1}{2}(D-2)(D-3)}\;,
\end{eqnarray}
\end{subequations}
where $c_s = \sqrt{\partial P/\partial \varepsilon}$
is the velocity of sound.
Here, we introduced the abbreviations
\begin{eqnarray*}
f & = & i\omega\, \tau_{\pi} +1\;, \qquad\;
f_{1}  =  -ik\, (\varepsilon+P)\;, \\
f_{2}  &=&  i\omega \, (\varepsilon+P)\;, \qquad
\Gamma  =  -ik\, \frac{2(D-2)}{D-1}\,\eta\;,\\
\Gamma_{1} & = & -ik\, \eta \;, \qquad\qquad
\Gamma_{2}  = ik \, \frac{2}{D-1}\, \eta \; .
\end{eqnarray*}
For nontrivial solutions of Eq.\ (\ref{lineq}),
the determinant of the matrix $A$
should vanish. This leads to the following conditions
for the dispersion relations $\omega(k)$:
\begin{subequations}
\begin{eqnarray}
f & = & 0 \label{eqn:rest_eq1}\;, \\
\det B = \left( \det B_0 \right)^{D-2}
& = & 0 \label{eqn:rest_eq2}\;, \\
\det T = \det\left(\begin{array}{ccc}
i\omega & f_1 & 0\\
-ik\,c_{s}^{2} & f_2 & -ik\\
0 & \Gamma & f
\end{array}\right) & = & 0 \label{eqn:rest_eq3}\;.
\end{eqnarray}
\end{subequations}

Equation (\ref{eqn:rest_eq1})
gives a purely imaginary frequency
\begin{equation}
\omega = \frac{i}{\tau_{\pi}}\;,
\end{equation}
which corresponds to a nonpropagating mode.
The degeneracy of this mode is $(D-3)[1 +(D-2)/2]$.

Equation (\ref{eqn:rest_eq2}) leads to a
complex frequency,
\begin{equation}
\omega=\frac{1}{2\tau_{\pi}}\left(
i \pm \sqrt{\frac{4\,\eta\,\tau_{\pi}}{\varepsilon+P}\,k^2-1}
\right)\;,
\end{equation}
corresponding to two propagating modes, if $k$ is
larger than the critical wavenumber
\begin{equation} \label{kc}
k_c = \sqrt{\frac{\varepsilon+P}{4 \, \eta\, \tau_\pi}}
\equiv \frac{\sqrt{b}}{2\,\tau_\pi}\;.
\end{equation}
Following Ref.\ \cite{baier}, we shall call these modes
shear modes. There are in total $2(D-2)$ shear modes.

Equation (\ref{eqn:rest_eq3}) gives the same
dispersion relation as Eq.\ (16) of Ref.\ \cite{dkkm3}, after
replacing $2(D-2)\eta/(D-1)$ with $\zeta_0$.
Introducing the sound attenuation length in $D$ space-time dimensions
\begin{equation}
\Gamma_s \equiv \frac{2(D-2)}{D-1}\, \frac{\eta}{\varepsilon +P}
\equiv \frac{2(D-2)}{D-1} \, \frac{\tau_\pi}{b}\;,
\end{equation}
the analytic solution in the limit of small wavenumber $k$ is
\begin{equation}
\omega = \left\{
\begin{array}{l} \displaystyle
\frac{i}{\tau_{\pi}}\;, \\
\displaystyle
\pm \, k\, c_s +
i\, \frac{\Gamma_s}{2}\, k^2\;,
\end{array}
\right.
\end{equation}
while for large wavenumber we obtain
\begin{equation} \label{lwn}
\omega = \left\{
\begin{array}{l} \displaystyle
\frac{i}{\tau_\pi}\, \left[ 1 +
\frac{\Gamma_s}{\tau_\pi c_s^2} \right]^{-1}\;,\\[0.3cm]
\displaystyle
\pm\, k \, c_s \sqrt{1+ \frac{\Gamma_s}{\tau_\pi c_s^2}}
+ \frac{i}{2 \tau_\pi}\,
\left[1+\frac{\tau_{\pi} c_s^2}{\Gamma_s}\right]^{-1}\;.
\end{array}
\right.
\end{equation}
This corresponds to another nonpropagating mode and two
propagating modes which we call sound modes
in accordance with Ref.\ \cite{baier}.
All imaginary parts are positive
and therefore the nonpropagating, as well as the
shear and sound modes are stable
around the hydrostatic equilibrium state.
This fact is already known from
the study of Hiscock and Lindblom \cite{his}.

\begin{figure}
\includegraphics[scale=1.0]{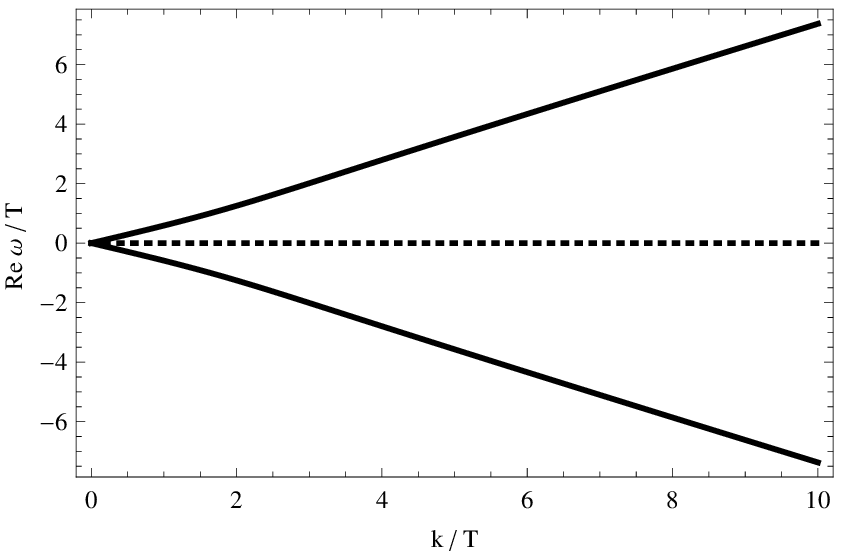}\includegraphics[scale=1.0]{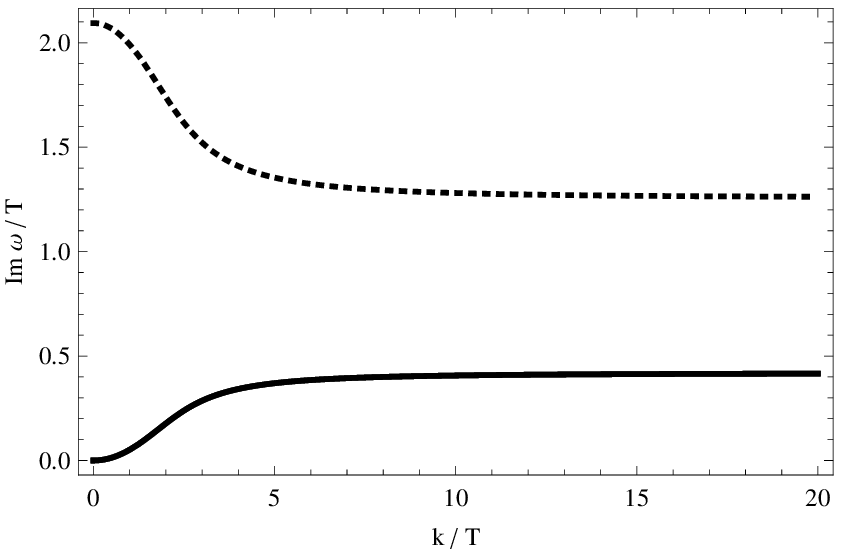}
\caption{The real parts (left panel) and the imaginary
parts (right panel) of the dispersion relations for
the sound modes (full lines) and the nonpropagating mode
(dashed line) obtained from Eq.\ (\ref{eqn:rest_eq3}).
The parameters are $a=\frac{1}{4\pi}\,,\;
b=6\,,\; c_{s}^{2}=\frac{1}{3}$ for the 3+1-dimensional
case, $D=4$.} \label{fig2}
\end{figure}

In order to discuss the issue of causality,
we follow Ref.\ \cite{his,dkkm3} and
study the group velocity defined as
\begin{equation}
v_g = \frac{\partial {\rm Re}\, \omega}{\partial k}\;.
\label{def_v}
\end{equation}
For the two nonpropagating modes, ${\rm Re}\, \omega=0$.
Consequently, in order to discuss causality, we
have to consider the behavior of the imaginary part \cite{dkkm3}.
Let us digress for the moment and consider
the diffusion equation with diffusion constant $D_0$.
There is a nonpropagating mode with
dispersion relation $\omega = iD_0 k^2$.
Moreover, it is known that the diffusion equation is acausal.
Therefore, we conjecture that a $k^2$ dependence of any
nonpropagating mode can be considered a sign of acausality.
In our case, the nonpropagating modes are either
independent of $k$, or have a weak $k$ dependence
(cf.\ Fig.\ \ref{fig2}).
According to our conjecture, we conclude
that the nonpropagating modes do not violate causality.

\begin{figure}
\includegraphics[scale=1.0]{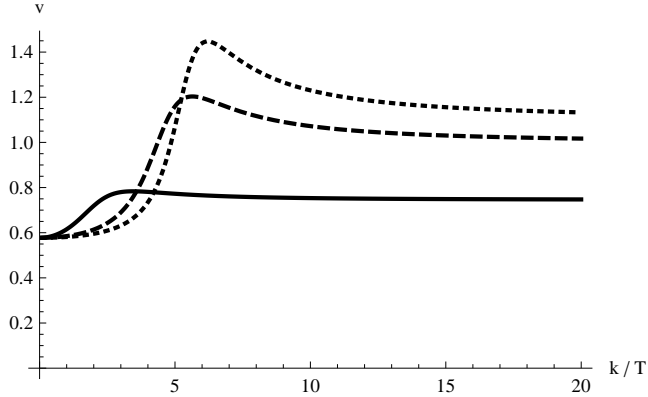}
\caption{The group velocity (\ref{vg}) for $a= 1/(4 \pi)\, , \;D=4\,
, \; c_{s}^{2}=\frac{1}{3}$, and $b=6$ (full line), $b=2$ (dashed
line), as well as $b=1.5$ (dotted line).}\label{v_sound}
\end{figure}

The dispersion relations resulting from Eq.\ (\ref{eqn:rest_eq3})
are shown in Fig.\ \ref{fig2}, and the corresponding group
velocity resulting from Eq.\ (\ref{def_v}) in Fig.\ref{v_sound}. The
group velocity has a maximum for a finite value of $k/T$ and
approaches its asymptotic value ($k \rightarrow \infty$) from above.
For small values of $b$, it may thus happen that the group velocity
becomes superluminal. Nevertheless, in Sec.\ \ref{causality} we
shall show that only the asymptotic value determines whether the
theory as a whole is causal or not. The asymptotic value of the
group velocity is
\begin{equation}
v_{g,{\rm sound}}^{\rm as} = \lim_{k\rightarrow \infty} \frac{\partial Re
\,\omega}{\partial k} = c_s\,
\sqrt{1+\frac{\Gamma_s}{\tau_{\pi}c_s^2 }}\; \label{asy_sol}.
\end{equation}
Consequently, for the asymptotic group velocity of sound waves to be
less than the speed of light,
$\tau_{\pi}$ and $\Gamma_s$ should satisfy the
following, so-called asymptotic causality condition:
\begin{equation}
\frac{\Gamma_s}{\tau_\pi} \le 1 - c_s^2\;\;
\Longleftrightarrow \;\;
\frac{1}{b} \equiv \frac{\eta}{\tau_{\pi}(\varepsilon + P)} \le
\frac{D-1}{2(D-2)}(1-c_s^2) \;. \label{CC}
\end{equation}
This is similar to the causality condition for the
group velocity in the case of bulk viscosity,
Eq.\ (21) of Ref.\ \cite{dkkm3}. For conformal fluids,
where $c_s^2 = 1/(D-1)$, the condition (\ref{CC}) simplifies
to $\Gamma_s \le (D-2)\tau_\pi/(D-1)$ or, equivalently, $b \geq 2$.
For example, for the values of $\eta$ and $\tau_\pi$
deduced from the AdS/CFT correspondence
\cite{baier,Heller:2007qt,push}, $\eta = s/(4 \pi)$,
$\tau_\pi = (2 - \ln 2)/(2 \pi T)$,
the condition (\ref{CC}) is always satisfied
because $b = 2 (2-\ln 2) \simeq 2.614 > 2$.

\begin{figure}
\includegraphics[scale=1.0]{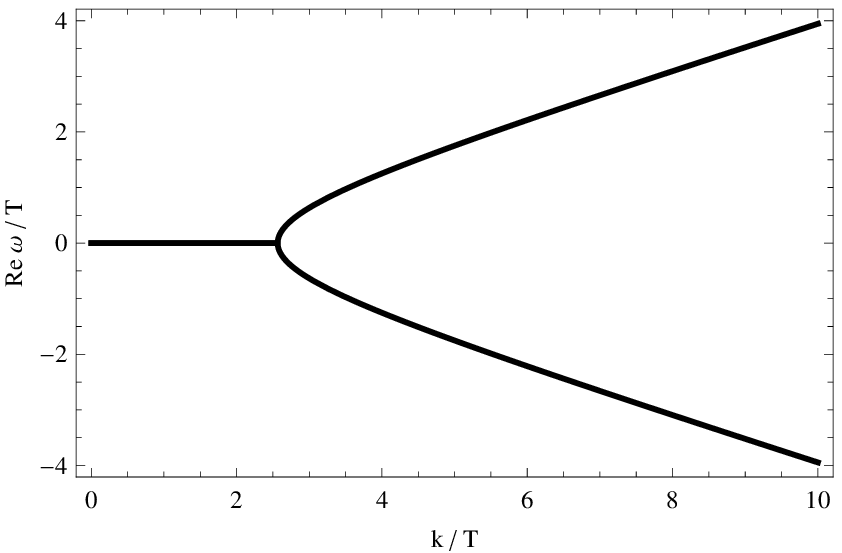}\includegraphics[scale=1.0]{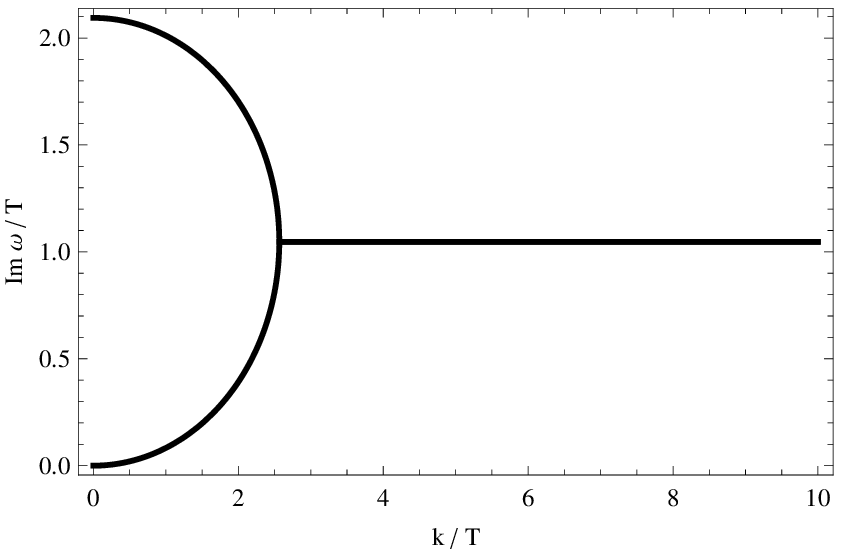}
\caption{The real parts (left panel) and the imaginary
parts (right panel) of the dispersion relations for
the shear modes obtained from Eq.\ (\ref{eqn:rest_eq2}).
The parameters are $a=\frac{1}{4\pi}\,,\;b=6\,,\;
c_{s}^{2}=\frac{1}{3}$ for the 3+1-dimensional case, $D=4$.}
\label{fig1}
\end{figure}

\begin{figure}
\includegraphics[scale=1.0]{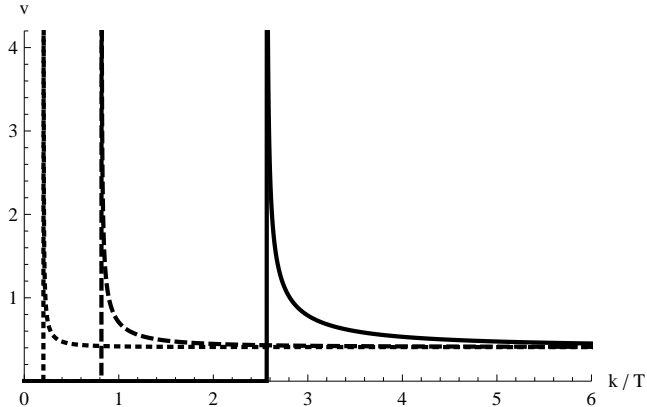}
\caption{The group velocity (\ref{vg}) for
$D=4\, , \; b=6\,,\; c_{s}^{2}=\frac{1}{3}$, and $a= 1/(4 \pi)$
(full line), $a=1/4$ (dashed line), as well as
$a=1$ (dotted line).} \label{fig3}
\end{figure}

The dispersion relations for the shear modes
resulting from Eq.\ (\ref{eqn:rest_eq2})
change their behavior from nonpropagating
to propagating at the critical wavenumber (\ref{kc}), as
shown in Fig.\ \ref{fig1}.
It should be noted that a similar behavior is observed
in the case of bulk viscosity,
cf.\ Fig.\ 1 in Ref.\ \cite{dkkm3}.
For wavenumbers larger than $k_c$,
the (modulus of the) group velocity of the propagating mode is
\begin{equation} \label{vg}
v_g = v_{g,{\rm shear}}^{\rm as}\,
\frac{k/k_c}{\sqrt{ (k/k_c)^2-1}}\;,
\end{equation}
where
\begin{equation} \label{asy_shear}
v_{g,{\rm shear}}^{\rm as} \equiv \frac{1}{\sqrt{2 \tau_\pi k_c}}
\equiv \sqrt{\frac{\eta}{\tau_\pi(\varepsilon
+P)}} \equiv \frac{1}{\sqrt{b}}
\end{equation}
is the asymptotic value of $v_g$ in the
large-wavenumber limit.
If the asymptotic causality condition (\ref{CC})
is satisfied, $v_{g,{\rm shear}}^{\rm as} \le \sqrt{(D-1)(1-c_s^2)/2(D-2)}$.
This is smaller than 1 for any value of $c_s$
and $D\ge 3$.
However, near the critical wavenumber $k_c$
the group velocity diverges, as shown in Fig.\ \ref{fig3}.
From the definitions of $k_c$, Eq.\ (\ref{kc}), and
the parameters $a,b$, Eqs.\ (\ref{param}), we
observe that $k_c/T = (2a \sqrt{b})^{-1}$. The
$1/a$-scaling of $k_c/T$ for fixed $b$ can be nicely observed
in Fig.\ \ref{fig3}.


In Sec.\ \ref{causality} we shall show that the apparent
violation of causality of the group velocity does not cause
the theory as a whole to become acausal. The important issue
is whether the asymptotic causality condition is fulfilled. If yes,
the theory is causal.

\begin{figure}
\includegraphics[scale=1.0]{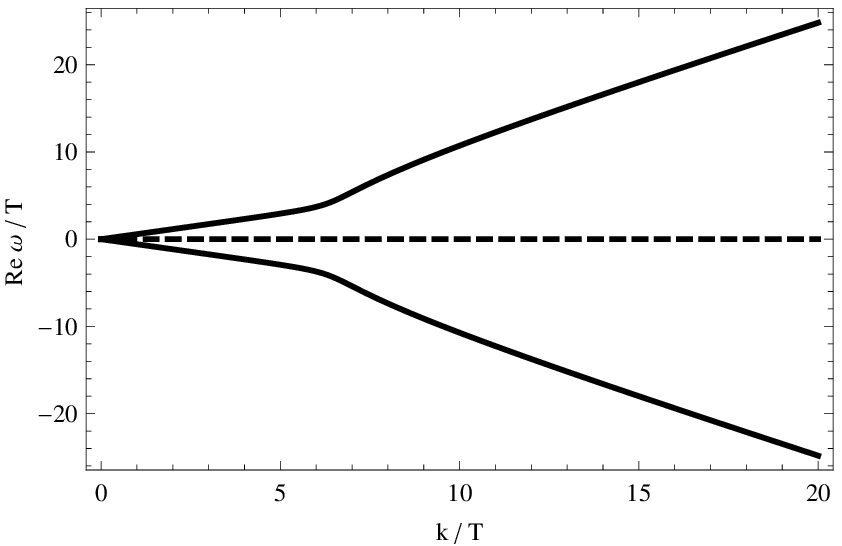}\includegraphics[scale=1.0]{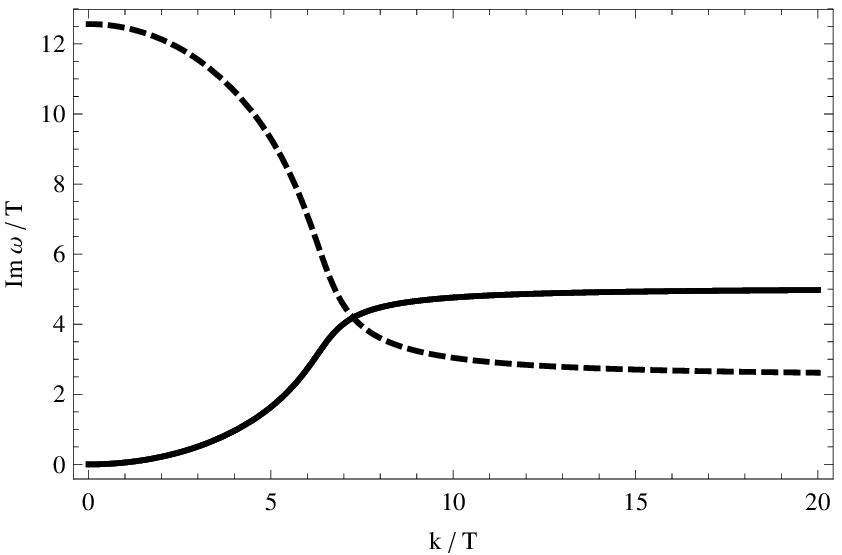}
\caption{The real parts (left panel) and the
imaginary parts (right panel) of the dispersion relations
for the sound modes obtained from
Eq.\ (\ref{eqn:rest_eq3}). The parameters are
$a=\frac{1}{4\pi}\,,\;b=1\,,\; c_{s}^{2}=\frac{1}{3}$ for the
3+1-dimensional case, $D=4$.} \label{ac_sound}
\end{figure}

\begin{figure}
\includegraphics[scale=1.0]{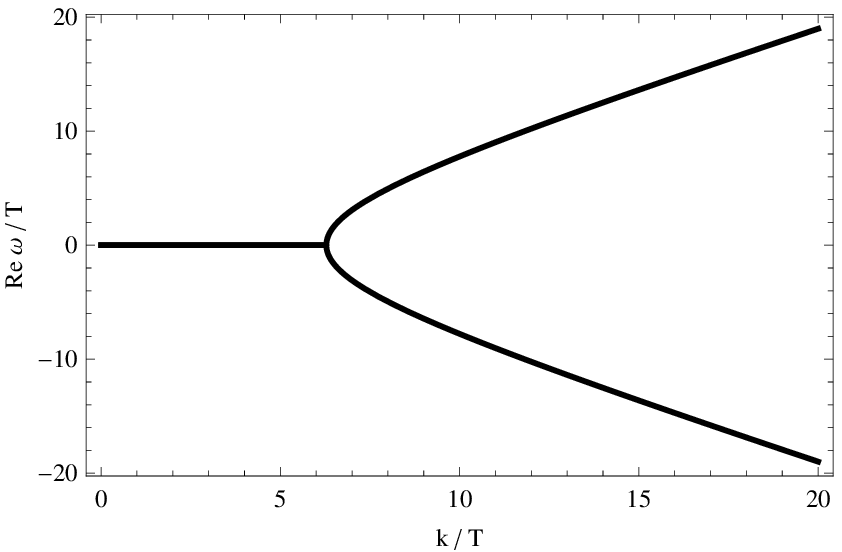}\includegraphics[scale=1.0]{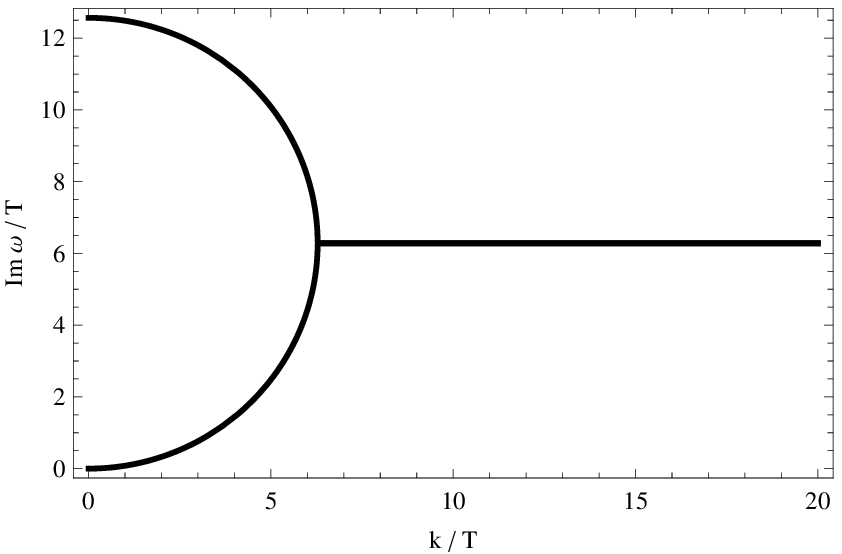}
\caption{The real parts (left panel) and the imaginary
parts (right panel) of the dispersion relations for the
shear modes obtained from Eq.\ (\ref{eqn:rest_eq2}).
The parameters are
$a=\frac{1}{4\pi}\,,\;b=1\,,\; c_{s}^{2}=\frac{1}{3}$ for the
3+1-dimensional case, $D=4$.} \label{ac_shear}
\end{figure}

We remark that, {\em in the local rest frame}, the {\em stability\/}
of the system of fluid-dynamical equations is {\em not\/} affected
if we choose a parameter set which violates the asymptotic
causality condition (\ref{CC}), for instance a conformal
fluid in $D=4$ dimensions and $b=1$. This is demonstrated for
the sound modes in Fig.\ \ref{ac_sound}, and for
the shear modes in Fig.\ \ref{ac_shear}.

\subsection{Competition of bulk and shear}

The question we would like to answer in this section is
whether the problem of the divergent group velocity
can be removed by adding bulk viscosity to the discussion.
For the sake of simplicity, we consider only the
2+1-dimensional case, i.e., $D=3$.
Similarly to Eqs.\ (\ref{param}), we introduce
the parametrization
\begin{equation}
\zeta=a_{1}s\;,\qquad
\tau_{\Pi}=\frac{\zeta}{\varepsilon+P}\,b_{1}\;.
\end{equation}
As before, the equations of motion (\ref{eqofmotion}) have
to be linearized, yielding Eq.\ (\ref{lineq}),
where now
\begin{equation}
X=(\delta \varepsilon,\delta u^{x}, \delta \pi^{xx},
\delta u^{y}, \delta \pi^{xy}, \delta \Pi)^{T}\;,
\end{equation}
and
\begin{eqnarray}
A & = & \left(\begin{array}{cccccc}
i\omega & -ik\,(\varepsilon+P) & 0 & 0 & 0 & 0\\
-ik\,c_{s}^{2} & i\omega\,(\varepsilon+P) & -ik & 0 & 0 & -ik\\
0 & -ik\, \eta & i \omega\, \tau_{\pi}+1 & 0 & 0 & 0\\
0 & 0 & 0 & i\omega\,(\varepsilon+P) & -ik & 0\\
0 & 0 & 0 & - ik\, \eta & i\omega\, \tau_{\pi}+1 & 0 \\
0 & -ik\, \zeta & 0 & 0 & 0 & i\omega\,\tau_{\Pi}+1
\end{array}\right)\;.
\end{eqnarray}
Then, the dispersion relations are given by solving the
following equations:
\begin{subequations}
\begin{eqnarray}
k^{2}\eta+i\omega\,(1 +i\omega \, \tau_{\pi})(\varepsilon+P)
&=& 0\;,\label{eqn:s+b_2}\\
i\omega k^{2}\, (1+i\omega \, \tau_{\Pi}) \,\eta
+(1+i\omega\, \tau_{\pi})
\left[i\omega k^{2}\, \zeta
+(1+i\omega\, \tau_{\Pi})(\varepsilon+P)
(c_{s}^{2}k^{2}-\omega^{2})\right]
&=& 0 \;.\label{eqn:s+b_1}
\end{eqnarray}
\end{subequations}
The dispersion relations resulting from sound and
bulk viscous modes, Eq.\ (\ref{eqn:s+b_1}),
are
\begin{equation}
\omega = \left\{ \begin{array}{l} \displaystyle
\frac{T}{2aa_1(b+b_1+bb_1c_s^2)} \left\{ \frac{}{}
ia(1+bc_s^2)+ia_1(1+b_1c_s^2) \right.
\\ \qquad  \left.\pm \left[ 4aa_1 c_s^2
(b+b_1+bb_1c_s^2)-(a+a_1+abc_s^2+a_1b_1c_s^2)^2 \right]^{1/2}
\right\}\;,\\
\displaystyle \pm k \sqrt{\frac{1}{b}
+ \frac{1}{b_{1}}+c_{s}^{2}}
+\frac{i\,T}{2(b+b_1+bb_1c_s^2)}\left(
\frac{b}{a_1b_1}+\frac{b_1}{ab} \right)\;,
\end{array}
\right.
\end{equation}
for large $k$, and
\begin{equation}
\omega= \left\{
\begin{array}{l} \displaystyle \frac{i}{\tau_\pi} \;,\\[0.3cm]
 \displaystyle \frac{i}{\tau_\Pi} \;,\\
\pm c_s^2k \; ,
\end{array}\right.
\end{equation} for small $k$.

Thus the asymptotic causality condition reads
\begin{equation}
\frac{1}{b_1} + \frac{1}{b}\equiv
\frac{\zeta}{\tau_{\Pi}(\varepsilon + P)} +
\frac{\eta}{\tau_{\pi}(\varepsilon + P)} \le 1-c^2_s \; .
\label{asy_bulk}
\end{equation}

On the other hand, the equation for the shear modes,
Eq.\ (\ref{eqn:s+b_2}), is the same as
Eq.\ (\ref{eqn:rest_eq2}) and hence the corresponding group
velocity again shows a divergence.
Thus, the inclusion of bulk
viscosity does not solve the problem of the divergent
group velocity.

\section{Stability in Lorentz-boosted frame} \label{III}

The discussion of causality and stability in the case
of nonzero bulk viscosity in a Lorentz-boosted frame
in Ref.\ \cite{dkkm3} has
shown that causality and stability are intimately related.
Relativistic dissipative fluid dynamics becomes unstable
if the group velocity exceeds the speed of light.
If this is still true in the case of nonzero
shear viscosity, the divergence of the group velocity found
in the rest frame may induce an instability in a moving frame.
In order to investigate this question,
we consider the stability of the hydrostatic state
observed from a Lorentz-boosted frame,
following Ref.\ \cite{dkkm3}. In this section, we restrict our
investigations to the case $D=4$.

We consider a frame moving with a velocity $\vec{V}$ with
respect to the hydrostatic state. Then, the total
fluid velocity $u^{ \prime \; \mu}$ is given by
\begin{equation}
u^{\prime \; \mu} = \left(
\begin{array}{cc}
\gamma_V & V \gamma_V \vec{n}^T \\
V \gamma_V \vec{n} & \gamma_V P_{\parallel}+Q_{\perp }
\end{array}
\right) u^{\mu},
\end{equation}
where $\gamma_V = 1/\sqrt{1-V^2}$,
$P_{\parallel }=\vec{n}\vec{n}^{T}$,
and $Q_{\perp}=1-P_{\parallel }$,
with $\vec{n}=\vec{V}/|\vec{V}|$.
We consider the two cases where the direction of the
Lorentz boost is parallel and where it is
perpendicular to the direction of propagation
of the perturbation; the latter we take to be the $x$ direction.

\subsection{Boost along the $x$ direction}

The perturbation of the fluid velocity is given by
\begin{equation}
u^{\prime \; \mu} = u^{\prime \; \mu}_0 +
\delta u^{\prime \; \mu} \; e^{i\omega t - ikx}\;,
\end{equation}
where
\begin{subequations}
\begin{eqnarray}
u^{\prime \; \mu}_0 &=& \gamma_V (1, V, 0, 0)\;, \\
\delta u^{\prime \; \mu} &=& (V\gamma_V \delta u^x,
\gamma_V \delta u^x, \delta u^y, \delta u^z) \;,
\end{eqnarray}
\end{subequations}
where $\delta u ^\mu$ is the velocity perturbation in the
local rest frame. The linearized fluid-dynamical equations
are again given by Eq.\ (\ref{lineq}), with
\begin{equation}
X = (\delta \varepsilon,\delta u^{x},\delta \pi^{xx},\delta
u^{y},\delta \pi^{xy},\delta u^{z},\delta \pi^{xz},\delta
\pi^{yy},\delta  \pi^{yz})^T \;,
\end{equation}
and
\begin{eqnarray}
A &=& \left(\begin{array}{cccc}
T_1 & 0 & 0 & 0\\
0 & B_1 & 0 & 0\\
G_1 & 0 & C_1 & 0\\
0 & 0 & 0 & E_1\end{array}\right)\; .
\end{eqnarray}
The submatrices are given by
\begin{subequations}
\begin{eqnarray}
T_1 &=&\gamma_{V}^2 \left(
\begin{array}{ccc}
i\omega(1+V^{2}c_{s}^{2})-ikV(1+c_{s}^{2}) & \;\; i[2\omega V
-k(1+V^{2})](\varepsilon+P) &
\;\; i\gamma_{V}^{-2}V(\omega V-k)\\
i\omega V(1+c_{s}^{2})-ik(V^{2} + c_{s}^{2}) &
\;\;
i\left[\omega(1+V^{2})-2kV\right](\varepsilon+P) & \;\;
i\gamma_V^{-2}(\omega V-k)\\
0 & \frac{4}{3}i\eta\gamma_{V}(\omega V-k) &\;\; \gamma_{V}^{-2}
F\end{array}
\right) \; ,\nonumber \\
\\
B_1 &=& {\rm diag}(B_{01},B_{01})\;,\qquad
B_{01}=\left(\begin{array}{cc} i\gamma_{V}(\omega-k
V)(\varepsilon+P) & \;\;
i(\omega V-k)\\
i\eta\gamma_{V}^{2}(\omega V-k) & \;\; F\end{array}\right)
\; , \\
G_1 &=& \left(\frac{}{} 0 \qquad
 -\frac{2}{3}i\eta\gamma_{V}(\omega V-k) \qquad 0 \right)  \;,\\
C_1 &=&  E_1 = F \;.
\end{eqnarray}
Here we abbreviated
\begin{equation}\label{F}
F=i\gamma_{V}(\omega-kV)\tau_{\pi}+1\;.
\end{equation}
\end{subequations}
Obviously,
\begin{eqnarray}
{\rm det} A = {\rm det} T_1 \times {\rm det} B_1 \times F^2 \;.
\end{eqnarray}
From $F^2=0$, we only obtain two trivial propagating modes
\begin{equation} \label{trivialmode}
\omega=\frac{i}{\gamma_{V}\tau_{\pi}}+kV\;.
\end{equation}
The group velocity is $ v_{g}=V$, which implies that these
modes correspond to the nonpropagating modes in the LRF.

From ${\rm det} B_1 =0$, we obtain
\begin{equation}
[iT+ab\gamma_{V}(kV-\omega)](kV-\omega)+a\gamma_{V}
(kV-\omega)^{2}T=0\; ,
\end{equation}
corresponding to the shear modes. There are in total four modes
satisfying this relation. The solutions are given by
\begin{equation}
\omega_{\pm}=\frac{1}{2a(b-V^{2})\gamma_{V}}\left[i\,
T-2a(1-b)kV\gamma_{V}\pm\sqrt{-T^2+4iakTV\gamma_{V}^{-1}
+4a^{2}bk^{2}\gamma_{V}^{-2}}\right] \; . \label{eq:solution01}
\end{equation}

On the other hand, the sound modes result from
\begin{eqnarray}\nonumber
\lefteqn{ c_s^2(\varepsilon+P )\left[ \frac{}{} 1  -  i \gamma_V
\tau_{\pi}  (k V-\omega )\right] \left\{ \frac{}{} k^2 \left[
\frac{}{} V^2+(V-1)^2 V \gamma_V
   ^2+1\right] \right.  } & &\\ \nonumber
& + & \left.  2 k V \omega  \left[\frac{}{} (V-1) V \gamma_V
^2-1\right]+ V^2 \omega   ^2-c_s^{-2}
(\omega -k V)^2 \frac{}{}\right\}\\[-0.1cm]
&+&\frac{4}{3} i \gamma_V \eta (k-V \omega )^2 \left\{ \frac{}{}
 k V
\left[\frac{}{} c_s^2 \gamma_V ^2 V (1-V)-1\right]+\omega
\right\}\qquad=0 \; .
\end{eqnarray}

\begin{figure}
\includegraphics[scale=1.0]{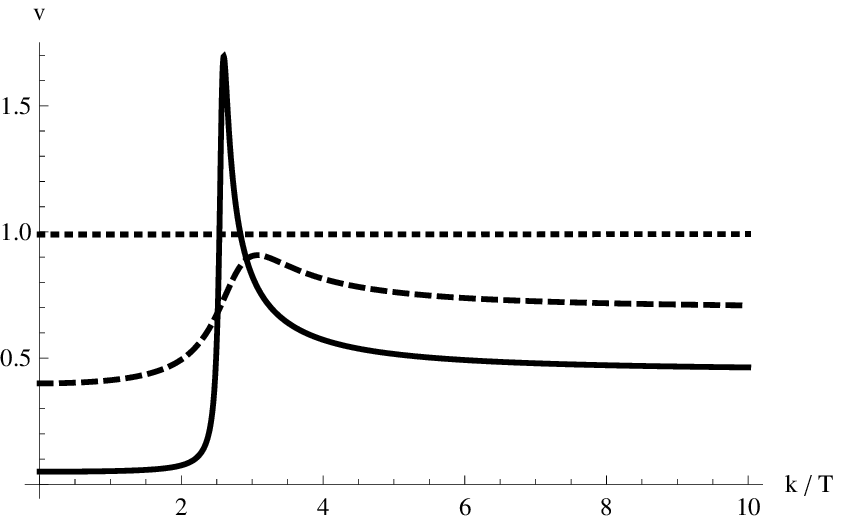}\includegraphics[scale=1.0]{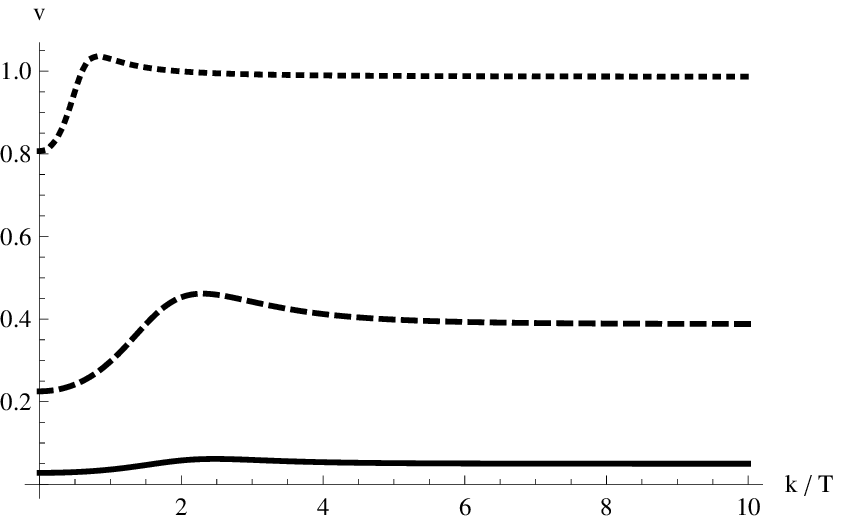}
\caption{The group velocity calculated for one of the
shear modes (left panel) and one of the sound modes
(right panel). We set $a=1/(4\pi),b=6,c_s^2=1/3$.
The solid line is for a boost velocity
$V=0.05$, the dashed line for $V=0.4$ and the
dotted line for $V=0.99$, respectively.} \label{fig4}
\end{figure}

In Fig.\ \ref{fig4}, the dependence of the group velocity
on the wavenumber is shown for various values of the
boost velocity $V$. The left panel
shows the behavior of one of the shear modes and the
right panel one of the sound modes. The parameter set used
here is $a=\frac{1}{4\pi},\,b=6,\,c_{s}^{2}=\frac{1}{3}$,
which satisfies the asymptotic causality
condition. We observe that the divergence of the group
velocity of the shear mode in the rest frame
is tempered by the Lorentz boost
to result in a peak of finite height. However, the
group velocity may still exceed the speed of light in
a certain range of wavenumbers. As we increase
the boost velocity, the peak height diminishes,
until the group velocity remains below the speed of
light for all wavenumbers. However, further increasing
the boost velocity leads to an acausal
group velocity in the {\em sound\/} mode.

\begin{figure}
\includegraphics[scale=1.0]{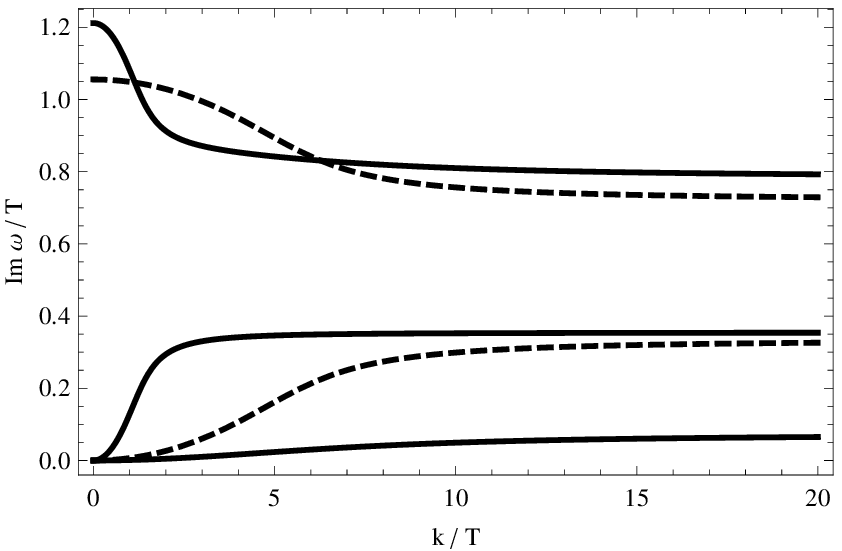}\includegraphics[scale=1.0]{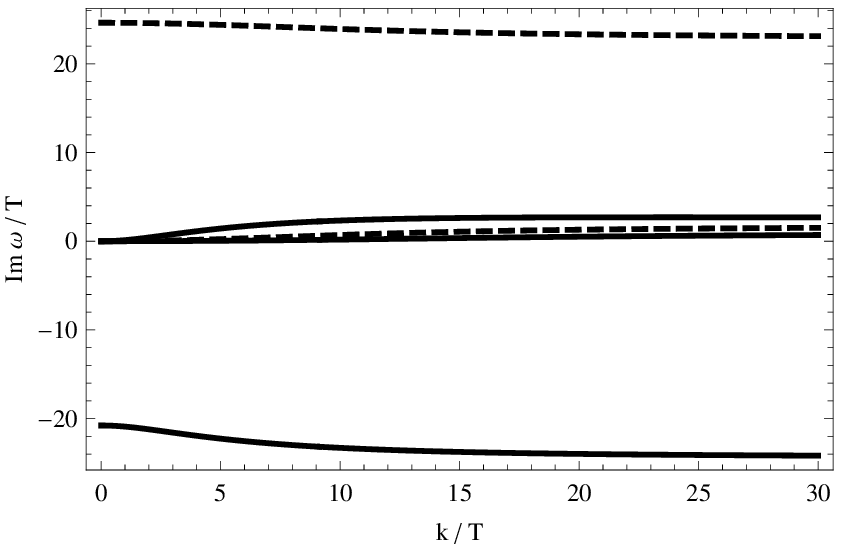}
\caption{The imaginary parts of the dispersion relations
for a boost in $x$ direction with velocity $V=0.9$.
The left panel shows the results for the parameter set
$a=\frac{1}{4\pi},\,b=6,\,c_{s}^{2}=\frac{1}{3}$, which
fulfills the asymptotic causality condition, while
the right panel is for
$a=\frac{1}{4\pi},\,b=1,\,c_{s}^{2}=\frac{1}{3}$,
which violates this condition.
The dashed lines are
for the shear modes, while the solid lines are for the
sound modes. } \label{fig5}
\end{figure}

Although the group velocity of the shear or the sound mode
may exceed the speed of light, as long as the asymptotic
causality condition is fulfilled, the theory is still
stable. This is demonstrated in the left panel of
Fig.\ \ref{fig5}, where the imaginary
parts of the modes are shown for the parameter set
$a=\frac{1}{4\pi},\,b=6,\,c_{s}^{2}=\frac{1}{3}$. We
observe that all imaginary parts are positive, indicating
the stability of the theory.

In contrast to the rest frame, where the theory is stable
even for parameters which violate the asymptotic
causality condition (\ref{CC}), this is no longer
the case in a Lorentz-boosted frame.
In the right panel of Fig.\ \ref{fig5}, the imaginary parts
of the modes are calculated with the parameter set
$a=\frac{1}{4\pi},\,b=1,\,c_{s}^{2}=\frac{1}{3}$. Now one
observes the appearance of negative imaginary parts,
indicating that the theory becomes unstable.

\subsection{Boost along the $y$ direction}

Now we consider a Lorentz boost along the $y$ direction.
The perturbation of the fluid velocity is given by
\begin{equation}
u^{\prime \; \mu} = u^{\prime \; \mu}_0
+ \delta u^{\prime \; \mu}\; e^{i\omega t - ikx} \;,
\end{equation}
where
\begin{subequations}
\begin{eqnarray}
u^{\prime \; \mu}_0 &=& \gamma_V (1, 0, V,0) \; , \\
\delta u^{\prime \; \mu} &=& (V\gamma_V \delta u^y, \delta u^x,
\gamma_V \delta u^y, \delta u^z ) \;.
\end{eqnarray}
\end{subequations}
Similarly to the preceding discussion, the linearized
fluid-dynamical equations take the form (\ref{lineq}), where
the matrix $A$ is
\begin{equation}
A=\left(
\begin{array}{cccc}  T_2 & H_1 & H_2 & 0 \\
H_3 & B_2 & H_4 & H_5\\
G_2 & H_6 & C_2 & 0 \\
0   & H_7 & 0   & E_2
\end{array}\right) \; ,
\end{equation}
with
\begin{subequations}
\begin{eqnarray}
T_2 &=& \left(\begin{array}{ccc}
i\omega\gamma_{V}^{2}(1+c_{s}^{2}V^2)
& -ik\gamma_{V} (\varepsilon+P) & 0 \\
-ikc_{s}^{2} & i\omega \gamma_{V}(\varepsilon+P) & -ik \\
0 & -\frac{4}{3}ik\eta & F_1
\end{array}\right) \; , \\
H_{1} & = & \left(\begin{array}{cccc}
2i\omega V(\varepsilon+P)\gamma_{V}^{2} & -ikV & 0 & 0\\
0 & i\omega V & 0 & 0\\
-\frac{2}{3}i\omega V\eta\gamma_{V} & 0 & 0 & 0\end{array}
\right)\;.\\
H_{2} & = & \left(\begin{array}{ccc}
i\omega V^{2} & 0 & 0\end{array}\right)^{T}\;,\\
H_{3} & = & \left(\begin{array}{ccc}
i\omega V\gamma_{V}^{2}(1+c_{s}^{2}) & -ikV\gamma_{V}
(\varepsilon+P) & 0\\
0 & i\omega V\gamma_{V}^{2}\eta & 0\\
0 & 0 & 0\\
0 & 0 & 0\end{array}\right)\;,\\
B_{2} & = & \left(\begin{array}{cccc}
i\omega\gamma_{V}^{2}(1+V^{2})(\varepsilon+P) & -ik & 0 & 0\\
-ik\gamma_{V}\eta & F_1 & 0 & 0\\
0 & 0 & i\omega\gamma_{V}(\varepsilon+P) & -ik\\
0 & 0 & -ik\eta & F_1\end{array}\right)\;,\\
H_{4} & = & \left(\begin{array}{cccc} i\omega V & 0 & 0 &
0\end{array}\right)^{T}\;,\qquad\qquad
H_{5}=\left(\begin{array}{cccc}
0 & 0 & i\omega V & 0\end{array}\right)^{T}\;,\\
G_{2} & = & \left(\begin{array}{ccc} 0 &
\frac{2}{3}ik\gamma_{V}^{2}\eta & 0\end{array}\right)\;,
\qquad\qquad
H_{6}=\left(\begin{array}{cccc}
\frac{4}{3}i\omega V\gamma_{V}^{3}\eta & 0 & 0 & 0
\end{array}\right)\;,\\
H_{7} & = & \left(\begin{array}{cccc} 0 & 0 & i\omega
V\gamma_{V}^{2}\eta & 0\end{array}\right)\;,\qquad\;\;\;
C_{2}=E_{2}=F_1\;.
\end{eqnarray}
\end{subequations}
Here we abbreviated
\begin{equation*}
F_1=i \omega \gamma_{V}\tau_{\pi}+1\;.
\end{equation*} The condition
${\rm det} A=0$ leads again to the following nine modes: three
nonpropagating modes, four shear modes and two sound modes.

The nonpropagating mode has almost the same form as that in
the LRF,
\begin{equation}
\omega=\frac{i}{\gamma_{V}\tau_{\pi}} \; .
\end{equation}
The shear modes
are given by the solution of the following equation
\begin{eqnarray}
k^{2}\eta+\gamma_{V}\omega\left[V^{2}\gamma_{V}\eta\omega
+(\varepsilon+P)(i-\gamma_{V}\tau_{\pi}\omega)\right]=0
\; ,
\end{eqnarray}
and the solutions are given by
\begin{equation}
\omega_{\pm}=\frac{1}{2a(b-V^{2})\gamma_{V}}\left[
i\,T\pm\sqrt{-T^2+4a^{2}bk^{2} -4a^{2}k^{2}V^{2}} \right] \; .
\end{equation}
We find that the critical wavenumber is now given by
$\tilde{k}_{c}=T/(2a\sqrt{b-V^{2}})$, below which
the shear modes become nonpropagating modes.

\begin{figure}[tbp]
\includegraphics[scale=0.85]{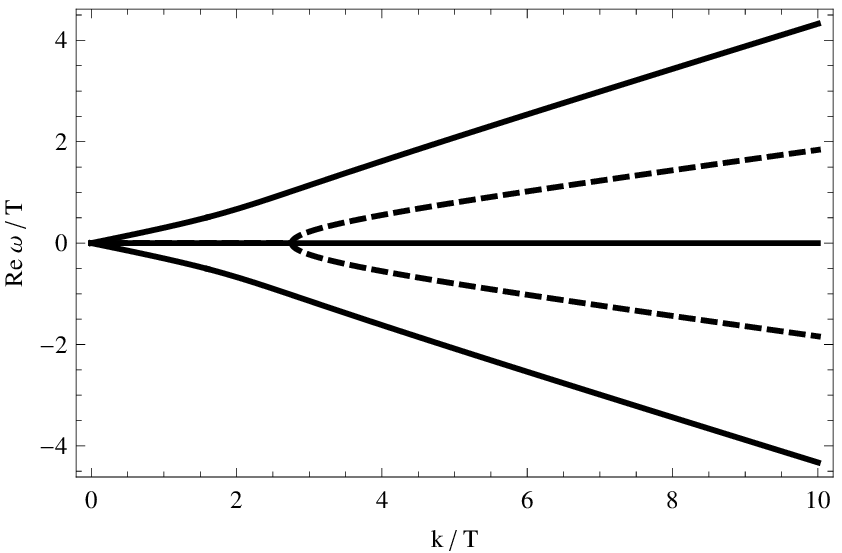}\includegraphics[scale=0.85]{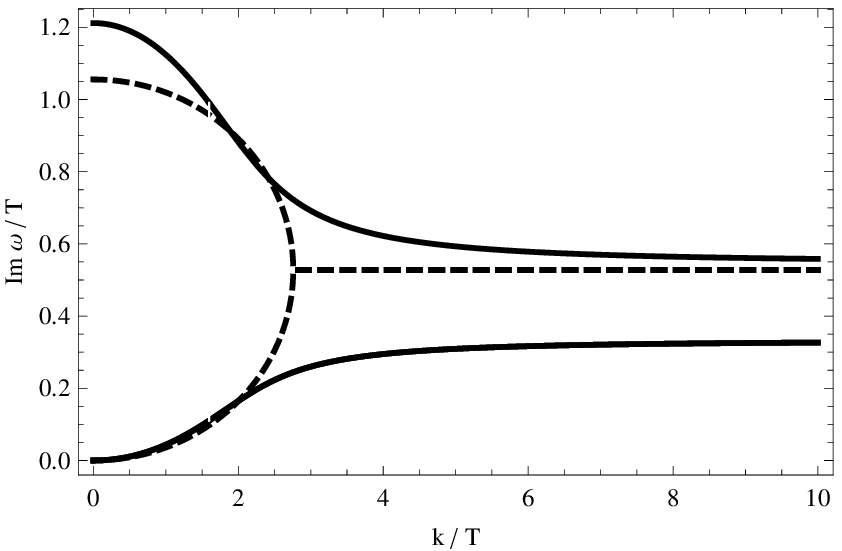}
\caption{The real and imaginary parts for the
dispersion relations of the shear modes (dashed lines)
and sound modes (solid lines), for a Lorentz boost in
$y$ direction. We use
$a=\frac{1}{4\pi},\,b=6,\,c_{s}^2=\frac{1}{3},\,V=0.9$
in the 3+1-dimensional case.
}
\label{fig8}
\end{figure}

On the other hand, the sound modes and another nonpropagating
mode result from
\begin{eqnarray}
\lefteqn{
3c_{s}^{2}(\varepsilon + P)
(-i+\gamma_{V}\tau_{\pi}\omega)
(k^{2}+V^{2}\gamma_{V}^{2}\omega^{2})}& & \nonumber \\
& + &\gamma_{V}\omega
\left\{4k^{2}\eta+\gamma_{V}\omega\left[3i(\varepsilon +P)
+4V^{2}\gamma_{V}\eta\omega-3(\varepsilon +P)
\gamma_V\tau_{\pi}\omega\right]\right\}=0
\; .
\end{eqnarray}
The real and imaginary parts of this dispersion relation
are calculated with a parameter set satisfying the
asymptotic causality condition. The results are
shown in Fig.\ \ref{fig8}.
One observes that the real parts are symmetric
around $\omega = 0$. This symmetry is due to
the fact that the direction of the
Lorentz boost is orthogonal to the direction of the
perturbation. The critical wave number $\tilde{k_c}$
where the shear mode changes from nonpropagating to
propagating mode can be clearly seen.
The imaginary parts are seen to be positive.
We confirmed that the imaginary parts become negative
if we use a parameter set which violates the asymptotic
causality condition.

\subsection{Causality of wave propagation} \label{causality}

In the preceding discussion we have seen that the theory is stable
if the asymptotic causality condition is fulfilled. The reverse
is in general not true, as the discussion in the local rest frame has shown,
since a stable theory may also violate the asymptotic causality condition.
However, the discussion in the Lorentz-boosted frame has revealed
that the stability of a theory is contingent upon whether
the {\em asymptotic\/} causality condition is fulfilled.

In this section, we shall show that the causality of
the theory as a whole is guaranteed if
the asymptotic stability condition is fulfilled.
The group velocity may become superluminal,
or even diverge, as long as this apparent
violation of causality is restricted to a finite range of momenta.
The argument leading to this conclusion is analogous to
that of Sommerfeld and Brillouin in classical electrodynamics
\cite{jackson,bri}. For instance, in the case of anomalous dispersion
the group velocity may become superluminal, but the causality of
the theory as a whole is not affected.

The change in a fluid-dynamical variable induced by a
general perturbation is given by
\begin{equation}
\delta X(x, t) =
\sum_{j} \int d\omega \, \widetilde{\delta X}_j (\omega)\,
e^{i\omega t -ik_j (\omega) x} \;, \label{B1}
\end{equation}
where $\delta X(x,t)$ stands for
$\delta \varepsilon$, $\delta u^{\mu}$, and
$\delta \pi^{\mu \nu}$.
The index $j$ denotes the different modes, i.e., the
shear modes, the sound modes etc.
The function $k_j (\omega)$ is the inverted
dispersion relation $\omega_j(k)$ of the respective mode.
The Fourier components are given by
\begin{equation}
\sum_j \widetilde{\delta X}_j(\omega)
= \frac{1}{2\pi} \int^{\infty}_{-\infty}
dt \, \delta X(0, t) \, e^{- i \omega t} \;.
\end{equation}

We assume that the incident wave has a well-defined front
that reaches $x=0$ not before $t=0$. Thus $\delta X(0,t) = 0 $
for $t<0$. This condition on $\delta X(0,t)$ ensures
that $\sum_j \widetilde{\delta X}_j(\omega)$
is analytic in the lower half of the complex $\omega$ plane
\cite{jackson}. On the other hand, in Sec.\ \ref{IIA} we have
found that the group velocity of the shear modes diverges
for certain values of $k$. These divergences correspond to
singularities in the complex $\omega$ plane.
However, if the asymptotic causality condition is
fulfilled, the imaginary part of the dispersion relation is
always positive, i.e., the singularities only appear in the
upper half of the complex $\omega$ plane. In this case,
the system is also stable.
On the other hand, if the asymptotic causality condition
is violated, the singularities may appear also
in the lower half-plane, i.e., for negative imaginary
part of the dispersion relation, and the system is unstable.

We shall now demonstrate that the divergences in the
group velocity do not violate
causality as long as the
asymptotic causality condition is satisfied, i.e., as long as the
asymptotic group velocity remains subluminal. To this end, we compute
Eq.\ (\ref{B1}) by contour integration in the complex
$\omega$ plane. To close the contour, we have to know
the asymptotic behavior of the dispersion relations.
In our calculation, we found that the real part of the
dispersion relation at large $k$
is proportional to $k$ [see Eq.\ (\ref{lwn})],
with a coefficient which is the large-$k$
limit of the group velocity, i.e., $v_{gj}^{\rm as}$,
\begin{equation}
\lim_{k \rightarrow \infty} {\rm Re}~\omega_j (k) = v_{gj}^{\rm as}\,
 k\; .
\end{equation}
Then, in the large-$k$ limit, the exponential becomes
\begin{equation}
\exp[i\omega t-ik_j(\omega)x]\rightarrow
\exp\left[-i\,\frac{\omega}{v_{gj}^{\rm as}}\,(x-v_{gj}^{\rm as}\,t)\right]
\;.
\end{equation}

In the case $x > v_{gj}^{\rm as}\,t $, we have to close the
integral contour in the lower half plane. If the asymptotic
causality condition is fulfilled,
there are no singularities in the lower half plane, and
Eq.\ (\ref{B1}) vanishes.
On the other hand, the contour should be closed
in the upper half plane if $x \le v_{gj}^{\rm as}\,t$.
Then, because of the singularities, Eq.\ (\ref{B1})
may have a nonzero value. However, as long as
we choose a parameter set for which the asymptotic
group velocity $v_{gj}^{\rm as}$
is smaller than the speed of light, i.e., for which
the asymptotic causality condition is fulfilled, the signal
propagation does not violate causality, since the locations
$x$ where the disturbance has travelled lie within the
cone given by $v_{gj}^{\rm as}$ which, in turn, lies within the lightcone,
q.e.d.

To conclude this section, we have shown that the asymptotic causality
condition not only implies stability in a general (Lorentz-boosted) frame,
but also causality of the theory as a whole.

\section{Characteristic velocities} \label{IV}

So far, we have analyzed the causality and stability of
relativistic dissipative fluid dynamics with shear
viscosity using a linear stability analysis.
However, there is another possibility to analyze
causality, namely by studying the characteristic
velocities. For the sake of simplicity, we consider
the 2+1-dimensional case with shear viscosity only.
The fluid-dynamical equations can be written in
the following form:
\begin{equation}
\left(A_{ab}^{t}\partial_{t}
+A_{ab}^{x}\partial_{x}
+A_{ab}^{y}\partial_{y} \right) Y_b=B_a\; , \label{eqn:non-lin}
\end{equation}
where $Y_b^T=(\varepsilon,u^{x},u^{y},\pi^{xx},\pi^{xy})$ and
$B_a^T=(0,\;0,\;0,\;\pi^{xx},\;\pi^{xy})$.
The expressions for the components of $A$
are given in the Appendix.
Then, as discussed in Ref.\ \cite{his}, the
characteristic velocities are defined as the roots of
the following equations,
\begin{subequations} \label{eq:new}
\begin{eqnarray}
\det(v_{x}A^{t}-A^{x}) & = & 0 \;  , \label{eq:newvx}\\
\det(v_{y}A^{t}-A^{y}) & = & 0 \;  . \label{eq:newvy}
\end{eqnarray}
\end{subequations}
For the case of bulk viscosity,
see Ref.\ \cite{dkkm3}.

For the sake of simplicity,
we consider $u^{\mu}=(1,\;0,\;0)$ and $\pi^{xx}=\pi^{xy}=0$.
Then, the characteristic velocities are given by
\begin{equation} \label{nonl_vel}
v_x=v_y= \left\{
\begin{array}{l}  0 \;,\\
\displaystyle \pm \sqrt{\frac{1}{b}} \;,\\
\displaystyle \pm  \sqrt{\frac{1}{b}+c_s^2} \; .
\end{array}\right.
\end{equation}
Interestingly, the second velocity is identical to the asymptotic group
velocity (\ref{asy_shear})
for the shear modes and the third velocity is the same as
the asymptotic group velocity (\ref{asy_sol}) for the sound modes
(since $D=3$).
As a matter of fact, if the asymptotic causality
condition (\ref{CC}) is satisfied,
the velocity (\ref{nonl_vel}) is smaller than the speed of light.

\begin{figure}
\includegraphics[scale=0.8]{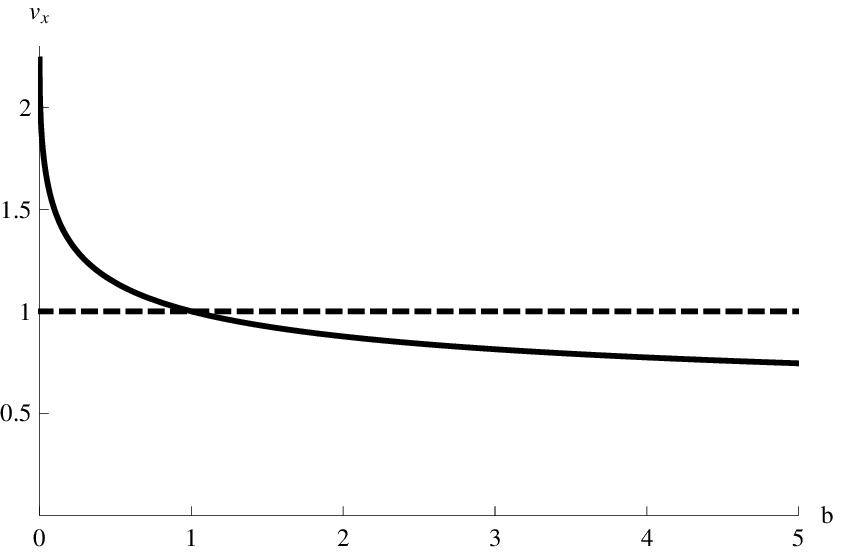}\includegraphics[scale=0.8]{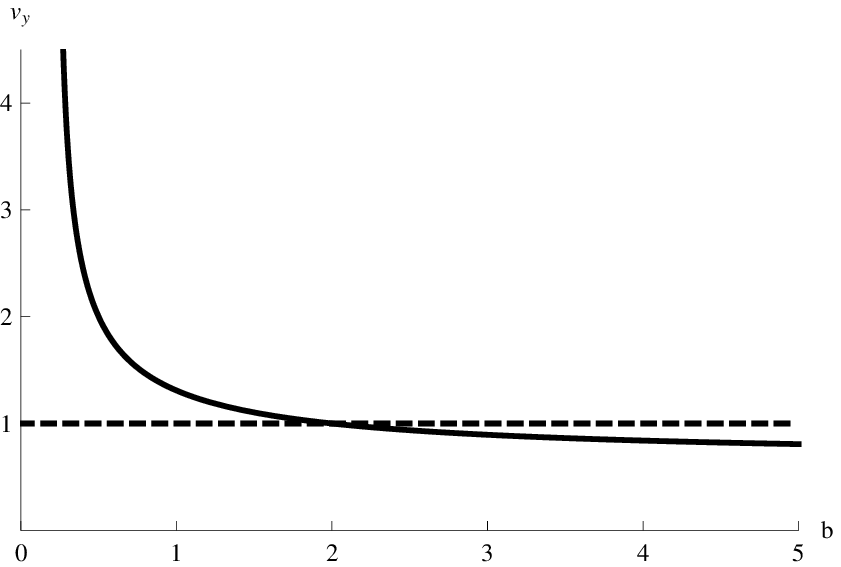}
\caption{One of the five characteristic velocities
determined from the roots of Eqs.\ (\ref{eq:new}).
The left panel is for $v_{x}$ and
the right panel is for $v_{y}$. We set
$u^\mu=(\sqrt{5}/2,\;1/2,\;0),\;\pi^{xx}=\pi^{xy}=0$, and
$c_s^2=1/2$.}\label{fig10}
\end{figure}

In Fig.\ \ref{fig10}, we show the $b$ dependence of
one of the five characteristic velocities.
We set
$u^\mu=(\sqrt{5}/2,\;1/2,\;0),\;\pi^{xx}=\pi^{xy}=0$, and
$c_s^2=1/2$.
The velocity exhibits a divergence at small values of $b$,
and thus exceeds the speed of light. This
divergence occurs also for at least one other
characteristic velocity.
As far as we have checked numerically,
in order to satisfy causality, one should use a
value of $b$ which is larger than about 2.
This condition is consistent with the asymptotic
causality condition (\ref{CC}).

\section{Concluding remarks} \label{V}

In this work, we have discussed the stability and causality
of relativistic dissipative fluid dynamics, based on a linear
stability analysis around a hydrostatic state.
Following the usual argument, we calculated the group
velocity from the dispersion relation of the perturbation.
We found that the group velocity diverges at a critical
wavenumber $k_c$. The appearance of the divergence is
independent of the dimensionality of space-time and
can be removed neither by tuning the
parameters of the theory nor by adding
bulk viscosity to the discussion.

Nevertheless, in the rest frame of the background
this acausal group velocity
does not cause the fluid to become unstable. Moreover,
investigating causality and stability in a Lorentz-boosted
frame, we found that the fluid-dynamical equations of
motion are stable,
if we choose parameters which satisfy a so-called
asymptotic causality condition. They become unstable if
this condition is violated.
In this sense, the problems of acausality and instability
are still correlated even in the case of shear viscosity,
as was already found for the case of bulk viscosity \cite{dkkm3}.

We have then demonstrated that the causality of the theory as
a whole is guaranteed if the asymptotic causality condition is
fulfilled. Therefore, a superluminal group velocity
in a finite range of momenta can cause the theory neither to
become acausal nor unstable.
Finally, we studied the characteristic velocities and found a
violation of causality for small values of $\tau_\pi(\varepsilon +P)/ \eta$,
but not for values which satisfy the asymptotic causality
condition.

The asymptotic causality condition
requires that the ratio $\tau_\pi/\Gamma_s$ is sufficiently
large, i.e., that the time scale $\tau_\pi$ over which the shear viscous
pressure relaxes towards its NS value is not too small compared
to the sound attenuation length $\Gamma_s \sim \eta/(\varepsilon +P)
\equiv \eta/(Ts)$.
This is an important finding for practitioners of fluid dynamics,
who frequently consider $\tau_\pi$
and the shear viscosity-to-entropy density ratio
$\eta/s$ to be independent from each other. We have demonstrated that
this is not the case if one wants the theory to remain causal. Therefore,
second-order theories of relativistic dissipative fluid dynamics
are not automatically causal by construction. Our
findings also illuminate why NS theory violates causality
from a different perspective, because
there $\tau_\pi \rightarrow 0$ while $\eta$ remains non-zero.

\section*{Acknowledgement}

Shi Pu thanks Zhe Xu and Qun Wang for helpful discussions.
We acknowledge G.\ Moore and the referee for valuable
comments concerning causality and the divergence of
the group velocity which have resulted in the discussion
presented in Sec.\ \ref{causality}.
This work was (financially) supported by the
Helmholtz International
Center for FAIR within the framework of the LOEWE program
(Landesoffensive zur Entwicklung Wissenschaftlich-\"Okonomischer
Exzellenz) launched by the State of Hesse.

\appendix

\section{Matrix elements
in Eq.\ (\ref{eqn:non-lin})}

The fluid-dynamical equations can be expressed in the form
(\ref{eqn:non-lin}).
Let us parameterize the velocity of the fluid as
$u^{\mu}=(\cosh\theta,\sinh\theta\cos\phi,\sinh\theta\sin\phi)$.
The matrix elements of $A_{ab}^{x}$ are
\begin{eqnarray*}
A^x_{11} & = &\left(c_s^2+1\right) \sinh \theta \cosh \theta \cos \phi\; ,\\
A^x_{12} & = & \frac{1}{2}\text{sech}^3\theta \left\{\frac{}{}2\sinh
^2\theta \left[(2 w+\pi^{xx}) \sin ^2\phi+3 w \cos ^2\phi-\pi^{xy}
\sin \phi
\cos \phi\right] \right. \\
& + & \left.  w \sinh ^4\theta (\cos (2 \phi )+3)+w+\pi^{xx}
\frac{}{} \right\}\; ,\\
A^x_{13} & = &\text{sech}^3\theta \left\{\frac{}{} \sinh ^2\theta
\cos \phi \left[(w-\pi^{xx}) \sin \phi
+\pi^{xy} \cos \phi\right]+w \sinh ^4\theta \sin \phi \cos \phi
+\pi^{xy}\right\}\; ,\\
A^x_{14} & = &\tanh \theta \cos \phi\; ,\\
A^x_{15} & = & \tanh \theta \sin \phi\; ,\\
A^x_{21} & = &\left(c_s^2+1\right) \sinh ^2\theta \cos ^2\phi+c_s^2\; ,\\
A^x_{22} & = &2 w \sinh \theta \cos \phi\; ,\\
A^x_{24} & = & A^x_{35}  = 1\; ,\\
A^x_{31} & = &\left(c_s^2+1\right) \sinh ^2\theta \sin \phi \cos \phi\; ,\\
A^x_{32} & = & w \sinh \theta \sin \phi\; ,\\
A^x_{33} & = & w \sinh \theta \cos \phi\; ,\\
A^x_{42} & = &\text{sech}^2\theta \left\{\frac{}{} \sinh ^4\theta
\cos ^2\phi \left[\eta +\tau_{\pi}\pi^{xx} \cos (2 \phi
)-\tau_{\pi}\pi^{xx}+\tau_{\pi}\pi^{xy}
\sin (2 \phi)\right] \right.\\
&+& \left. \sinh ^2\theta \left[2 (\eta -\tau_{\pi} \pi^{xx}) \cos
^2\phi
+\eta  \sin ^2\phi\right]+\eta \frac{}{} \right\}\; ,\\
A^x_{43} & = &-2 \tau_{\pi} \tanh ^2\theta \cos ^2\phi \left[\sinh
^2\theta
\cos \phi (\pi^{xy} \cos \phi-\pi^{xx} \sin \phi)+\pi^{xy}\right]\; ,\\
A^x_{44} & = & A^x_{55} = \tau _ {\pi } \sinh \theta \cos \phi\; ,
\end{eqnarray*}
\begin{eqnarray*}
A^x_{52} & = &\frac{\tanh ^2\theta \cos \phi}{2(\sinh ^2\theta \cos
^2\phi+1)}\left\{\frac{}{}-2\sinh ^2\theta \left(\pi^{xx} \sin
^3\phi +2 \pi^{xx} \sin \phi \cos ^2\phi+\pi^{xy} \cos ^3\phi\right)
\right. \\
&+& \left.  \sinh ^4\theta \sin ^2(2 \phi ) (\pi^{xy} \cos \phi
-2\pi^{xx} \sin \phi)-2\pi^{xx} \sin \phi-2\pi^{xy} \cos \phi
\frac{}{} \right\}
\; ,\\
A^x_{53} & = & \frac{1}{2}\text{sech}^2\theta \left\{\frac{}{}2\sinh
^4\theta \cos ^2\phi \left[\eta -\tau_{\pi}\pi^{xx} \cos (2 \phi)+
\tau_{\pi}\pi^{xx}-\tau_{\pi}\pi^{xy} \sin (2 \phi)\right]\right. \\
 &+& \left.  \sinh ^2\theta \left[(\eta +\tau_{\pi}\pi^{xx}) \cos (2\phi )
+3 \eta +\tau_{\pi}\pi^{xx}-\tau_{\pi}\pi^{xy} \sin (2 \phi
)\right]+2\eta \frac{}{} \right\}\; .
\end{eqnarray*}
The matrix elements of $A_{ab}^{t}$  are given by
\begin{eqnarray*}
A^t_{11}& = &\frac{1}{2} \left[\left(c_s^2+1\right) \cosh (2 \theta
)-c_s^2+1\right]\; , \\
A^t_{12}& = &\frac{2 \sinh \theta}{\left(\sinh ^2\theta \cos
^2\phi+1\right)^2}\left\{ \frac{}{} \sinh ^2\theta \cos \phi \left(2
w \cos ^2\phi+\pi^{xx} \sin ^2\phi-\pi^{xy} \sin \phi \cos
\phi\right)\right.\\
 &+& \left. w \sinh ^4\theta \cos ^5\phi+(w+\pi^{xx}) \cos
\phi+\pi^{xy} \sin \phi \frac{}{} \right\}
\; , \\
A^t_{13}& = &2 \sinh \theta \left(w \sin \phi+\frac{\pi^{xy} \cos \phi
-\pi^{xx} \sin \phi}{\sinh ^2\theta \cos ^2\phi+1}\right)\; ,\\
A^t_{14}& = &\frac{\cos (2 \phi )}{\text{csch}^2\theta+\cos
^2\phi}\; ,\\
A^t_{15}& = &\frac{\sin (2 \phi )}{\text{csch}^2\theta+\cos ^2\phi}\; ,\\
A^t_{21}& = &\left(c_s^2+1\right) \sinh \theta \cosh \theta \cos
\phi\; ,\\
A^t_{31} & = & \left(c_s^2+1\right) \sinh \theta \cosh \theta \sin \phi\; ,\\
A^t_{22}& = & \frac{\text{sech}^3 \theta}{2} \left\{\frac{}{}2\sinh
^2\theta \left[(2 w+ \pi^{xx}) \sin ^2\phi+3 w \cos ^2\phi-\pi^{xy}
\sin \phi \cos
\phi\right] \right. \\
&+& \left.  w \sinh ^4\theta \left[\cos (2 \phi )+3\right]+2w
+2\pi^{xx} \frac{}{}\right\}\; ,\\
A^t_{23}& = &\text{sech}^3\theta \left\{\sinh ^2\theta \cos \phi
\left[w \sinh ^2\theta \sin \phi
+(w-\pi^{xx}) \sin \phi+\pi^{xy} \cos \phi\right]+\pi^{xy}\right\}\; ,\\
A^t_{24}& = &\tanh \theta \cos \phi\; , \\
A^t_{25}&  = & \tanh \theta \sin \phi\; ,\\
A^t_{32}& = &\frac{\text{sech}^3\theta}{\left(\sinh ^2\theta \cos
^2\phi+1\right)^2} \left\{ \frac{}{}\sinh ^2\theta \left[(w+3
\pi^{xx})
 \sin \phi \cos \phi+3 \pi^{xy} \sin ^2\phi+2 \pi^{xy} \cos
 ^2\phi\right] \right. \\
 &+& \left.\sinh ^4\theta \left[3 (w+\pi^{xx}) \sin \phi \cos ^3\phi
 +(w+5 \pi^{xx}) \sin ^3\phi \cos \phi+2 \pi^{xy} \sin ^4\phi
 +\pi^{xy} \cos ^4\phi\right] \right. \\
 &+& \left. \frac{1}{16} \sinh ^6\theta \left[10 \sin (2 \phi )
 +\sin (4 \phi )\right] \left[(w-\pi^{xx}) \cos (2 \phi )+w+\pi^{xx}
-\pi^{xy} \sin (2 \phi
 )\right] \right.\\
 &+& \left. w \sinh ^8\theta \sin \phi \cos ^5\phi+\pi^{xy} \frac{}{} \right\}
 \; ,
\end{eqnarray*}
\begin{eqnarray*}
A^t_{33}& = &\frac{\text{sech}^3\theta }{8 \left(\sinh ^2\theta \cos
^2\phi+1\right)}\left\{\frac{}{} \sinh ^4\theta \left[4 (w+2
\pi^{xx}) \cos (2
\phi )+(\pi^{xx}-w) \cos (4 \phi )+21 w \right.\right. \\
&-&\left.\left. 9 \pi^{xx} +10 \pi^{xy} \sin (2 \phi )+\pi^{xy} \sin
(4 \phi )\right] +4 \sinh ^2\theta \left[6 w+2 \pi^{xx} \cos (2 \phi
)-4 \pi^{xx} \right. \right. \\
&+& \left. \left.3 \pi^{xy} \sin (2 \phi )\right]-4 w \sinh ^6\theta
\cos ^2\phi \left[\cos (2 \phi )-3\right]
+8 w-8 \pi^{xx} \frac{}{} \right\}\; ,\\
A^t_{34}& = &-\frac{\tanh \theta \sin \phi \left(\sinh ^2\theta
\sin ^2\phi+1\right)}{\sinh ^2\theta \cos ^2\phi+1}\; ,\\
A^t_{35}& = &\frac{\tanh \theta \cos \phi }{2 \sinh ^2\theta \cos
^2\phi+2} \left\{\frac{}{} 2-\sinh ^2\theta
[ \cos (2 \phi )-3 ]\right\} \; ,\\
A^t_{42}& = &\tanh \theta \cos \phi \left\{\frac{}{}\sinh ^2\theta
\left\{2 \sin \phi \left[(\eta -\tau_{\pi}\pi^{xx}) \sin
\phi+\tau_{\pi}\pi^{xy} \cos \phi\right]
+\eta  \cos ^2\phi\right\}+\eta -2 \tau_{\pi}\pi^{xx}\right\}\; ,\\
A^t_{43}& = & - \tanh \theta \left\{\frac{}{} \sinh ^2\theta \cos
^2\phi \left[(\eta -2 \tau_{\pi} \pi^{xx}) \sin \phi+2 \tau_{\pi}
\pi^{xy} \cos \phi\right]+\eta \sin \phi
+2 \tau_{\pi} \pi^{xy} \cos \phi\right\}\; ,\\
A^t_{44}& = & A^t_{55}= \tau _ {\pi } \cosh \theta\; ,\\
A^t_{52}& = &\frac{\tanh \theta}{4 \sinh ^2\theta \cos ^2\phi+4}
\left\{\frac{}{} -2 \sinh ^2\theta \left\{\sin \phi \left[-2 \eta +
\tau_{\pi} \pi^{xx} \cos (2 \phi )+3 \tau_{\pi}\pi^{xx}\right]+2
 \tau_{\pi} \pi^{xy} \cos
^3\phi\right\} \right. \\
&+& \left. \sinh ^4\theta \sin ^2(2 \phi ) \left[(\eta -2 \tau_{\pi}
\pi^{xx}) \sin \phi+2 \tau_{\pi} \pi^{xy} \cos \phi\right] +4 (\eta
- \tau_{\pi} \pi^{xx}) \sin \phi-4 \tau_{\pi} \pi^{xy} \cos \phi
\frac{}{} \right\}
\; ,\\
A^t_{53}& = &\tanh \theta \left\{ \frac{}{}\sinh ^2\theta \left[\eta
\cos ^3\phi +\tau_{\pi} \pi^{xx} \sin \phi \sin (2 \phi )-2
\tau_{\pi} \pi^{xy} \sin \phi
\cos^2\phi\right] \right. \\
&+&\left.(\eta + \tau_{\pi} \pi^{xx}) \cos \phi- \tau_{\pi} \pi^{xy}
\sin \phi \frac{}{}\right\}\; .
\end{eqnarray*}
The matrix elements of $A_{ab}^{y}$ are
\begin{eqnarray*}
A^y_{11}&=&\left(c_s^2+1\right) \sinh \theta \cosh \theta \sin
\phi\; , \\
A^y_{21} & =& \left(c_s^2+1\right) \sinh ^2\theta \sin \phi \cos \phi\; ,\\
A^y_{12}&=&\frac{\text{sech}^3\theta }{\left(\sinh ^2\theta \cos
^2\phi+1\right)^2} \left\{\frac{}{}\sinh ^2\theta \left[(w+3
\pi^{xx}) \sin \phi \cos \phi+3 \pi^{xy} \sin ^2\phi+2 \pi^{xy} \cos
^2\phi\right]\right. \\
 &+& \left. \sinh ^4\theta \left[3 (w+\pi^{xx}) \sin
\phi \cos ^3\phi +(w+5 \pi^{xx}) \sin ^3\phi \cos \phi+2 \pi^{xy}
\sin ^4\phi +\pi^{xy} \cos ^4\phi\right] \right. \\
&+&\left. \frac{1}{16} \sinh ^6\theta [10 \sin (2 \phi ) +\sin (4
\phi )] [(w-\pi^{xx}) \cos (2 \phi )+w+\pi^{xx}-\pi^{xy} \sin (2
\phi )] \right. \\
 &+& \left. w \sinh ^8\theta \sin \phi \cos
^5\phi+\pi^{xy}\frac{}{} \right\}
\; ,\\
A^y_{13}&=&\frac{\text{sech}^3\theta}{8 \left(\sinh ^2\theta \cos
^2\phi+1\right)} \left\{\frac{}{}\sinh ^4\theta [4 (w+2 \pi^{xx})
 \cos (2 \phi )+(\pi^{xx}-w) \cos (4 \phi )+21 w \right.\\
 &-&\left. 9 \pi^{xx}
 +10 \pi^{xy} \sin (2 \phi )+\pi^{xy} \sin (4 \phi )]
 +4 \sinh ^2\theta [6 w+2 \pi^{xx} \cos (2 \phi )-4 \pi^{xx} \right. \\
 &+& \left. 3 \pi^{xy} \sin (2 \phi
 )]
 -4 w \sinh ^6\theta \cos ^2\phi [\cos (2 \phi )-3]+8 w-
8 \pi^{xx}\frac{}{}\right\}
 \; ,\\
A^y_{14}&=&-\frac{\tanh \theta \sin \phi \left(\sinh ^2\theta
\sin ^2\phi+1\right)}{\sinh ^2\theta \cos ^2\phi+1}\; ,
\end{eqnarray*}
\begin{eqnarray*}
A^y_{15}&=&\frac{\tanh \theta \cos \phi }{2 \sinh ^2\theta \cos
^2\phi+2}
\left\{ \frac{}{} 2-\sinh ^2\theta \left[\cos (2 \phi )-3\right]\right\}\; ,\\
A^y_{22}&=&w \sinh \theta \sin \phi\;, \\
A^y_{23}& = & w \sinh \theta \cos \phi\; ,\\
A^y_{25}&=&1\;,\\
A^y_{31}& =& \left(c_s^2+1\right) \sinh ^2\theta \sin ^2\phi+c_s^2\; ,\\
A^y_{32}&=&\frac{2 \sinh \theta \left[\sinh ^2\theta \sin \phi \cos
\phi (\pi^{xx} \sin \phi-\pi^{xy} \cos \phi)+\pi^{xx} \cos \phi
+\pi^{xy} \sin \phi\right]}{\left(\sinh ^2\theta \cos ^2\phi+1\right)^2}\; ,\\
A^y_{33}&=&2 \sinh \theta \left(w \sin \phi+\frac{\pi^{xy} \cos \phi
-\pi^{xx} \sin \phi}{\sinh ^2\theta \cos ^2\phi+1}\right)\; ,\\
A^y_{34}&=&-\frac{\sinh ^2\theta \sin ^2\phi+1}{\sinh ^2\theta \cos
^2\phi+1}\; ,\\
A^y_{35}&=&\frac{\sin (2 \phi )}{\text{csch}^2\theta+\cos ^2\phi}\; ,\\
A^y_{42}&=& \tanh ^2\theta \sin \phi \cos \phi \left\{\frac{}{}
\sinh ^2\theta [2 \eta +\tau_{\pi}\pi^{xx} \cos (2 \phi
)-\tau_{\pi}\pi^{xx}
+\tau_{\pi}\pi^{xy} \sin (2 \phi )]+2\eta -2\tau_{\pi}\pi^{xx}\right\}\; ,\\
A^y_{43}&=&-\frac{ \text{sech}^2\theta}{2} \left\{\frac{}{} 2 \sinh
^4\theta \cos ^2\phi [\eta +\tau_{\pi}\pi^{xx} \cos (2 \phi
)-\tau_{\pi}\pi^{xx}+\tau_{\pi}\pi^{xy} \sin (2
\phi )] \right. \\
 &+& \left. \sinh ^2\theta \left\{\eta  [\cos (2 \phi )+3]+2\tau_{\pi}
\pi^{xy} \sin (2 \phi )\right\}
+2 \eta \frac{}{}\right\}\; ,\\
A^y_{44}&=& A^y_{55} = \tau _ {\pi } \sinh \theta \sin \phi\; ,\\
A^y_{52}&=&\frac{\tanh ^2\theta}{8 (\sinh ^2\theta \cos
^2\phi+1)}\left\{\frac{}{} \sinh ^2\theta
\left[(\tau_{\pi}\pi^{xx}-\eta ) \cos (4 \phi
)+9 \eta +4 \tau_{\pi}\pi^{xx} \cos (2 \phi ) - 5 \tau_{\pi}\pi^{xx}\right.\right. \\
&-& \left.\left. 8 \tau_{\pi}\pi^{xy} \sin \phi \cos ^3\phi\right]+2
\sinh ^4\theta \sin ^2(2 \phi ) [\eta +\tau_{\pi}\pi^{xx} \cos (2
\phi
)-\tau_{\pi}\pi^{xx}+\tau_{\pi}\pi^{xy} \sin (2 \phi )] \right. \\
 &+& \left.  4 [4 \eta
+\tau_{\pi}\pi^{xx} \cos (2 \phi
)-\tau_{\pi}\pi^{xx}-\tau_{\pi}\pi^{xy} \sin (2 \phi )]
+8 \eta  \text{csch}^2\theta \frac{}{}\right\}\; ,\\
A^y_{53}&=&\tau_{\pi}\tanh ^2\theta \sin \phi \left[\sinh ^2\theta
\sin (2 \phi ) (\pi^{xx} \sin \phi-\pi^{xy} \cos \phi) +\pi^{xx}
\cos \phi-\pi^{xy} \sin \phi\right]\; ,
\end{eqnarray*}
where we defined $w=\varepsilon+P$. All other elements
vanish.


\begin{thebibliography}{1}


\bibitem{Adams:2003zg}
  J.~Adams {\it et al.}  [STAR Collaboration],
  Phys.\ Rev.\ Lett.\  {\bf 92}, 062301 (2004)
  [arXiv:nucl-ex/0310029].

\bibitem{Adams:2003am}
  J.~Adams {\it et al.}  [STAR Collaboration],
  Phys.\ Rev.\ Lett.\  {\bf 92}, 052302 (2004)
  [arXiv:nucl-ex/0306007].

\bibitem{Sorensen:2003kp}
  P.~R.~Sorensen,
  arXiv:nucl-ex/0309003.

\bibitem{Adler:2002pu}
  C.~Adler {\it et al.}  [STAR Collaboration],
  Phys.\ Rev.\  C {\bf 66}, 034904 (2002)
  [arXiv:nucl-ex/0206001].


\bibitem{review}
See for example, P. Huovinen and P.V. Ruuskanen, Ann.
Rev. Nucl. Part. Sci. \textbf{56}, 163 (2006); Y. Hama, T. Kodama and O.
Socolowski, Braz. J. Phys. \textbf{35}, 24 (2005), Jean-Yves Ollitrault,
Euro. J. Phys. \textbf{29}, 275 (2008) and references therein.








\bibitem{press}
\emph{RHIC Scientists Serve Up "Perfect" Liquid},\\
{\scriptsize{\url{http://www.bnl.gov/bnlweb/pubaf/pr/PR_display.asp?prID=05-38}}}

\bibitem{Gyulassy:2004zy}
  M.~Gyulassy and L.~McLerran,
  Nucl.\ Phys.\  A {\bf 750}, 30 (2005)
  [arXiv:nucl-th/0405013].

\bibitem{Shuryak:2004cy}
  E.~V.~Shuryak,
  Nucl.\ Phys.\  A {\bf 750}, 64 (2005)
  [arXiv:hep-ph/0405066].


\bibitem{Danielewicz:1984ww}
  P.~Danielewicz and M.~Gyulassy,
  Phys.\ Rev.\  D {\bf 31}, 53 (1985).

\bibitem{Kovtun:2004de}
  P.~Kovtun, D.~T.~Son and A.~O.~Starinets,
  Phys.\ Rev.\ Lett.\  {\bf 94}, 111601 (2005)
  [arXiv:hep-th/0405231].

\bibitem{LL}
L.~D.~Landau, E.~M.~Lifshitz: \emph{Fluid Mechanics}, (Pergamon
Press, New York, 1959), Sections 133-136




\bibitem{muronga}
A. Muronga, Phys. Rev. Lett. {\bf 88}, 062302 (2002)
[Erratum ibid. {\bf 89}, 159901 (2002)].

\bibitem{roma}
P. Romatschke and U. Romatschke, Phys. Rev. Lett.
\textbf{99}, 172301 (2007).


\bibitem{luzum1}
M. Luzum and P. Romatschke,
Phys.\ Rev.\ Lett.\  {\bf 103}, 262302 (2009)
[arXiv:0901.4588 [nucl-th]].


\bibitem{luzum2}
M. Luzum and P. Romatschke,
Phys.\ Rev.\  C {\bf 78}, 034915 (2008)
[Erratum-ibid.\  C {\bf 79}, 039903 (2009)]
[arXiv:0804.4015 [nucl-th]].

\bibitem{Song:2007ux}
H.~Song and U.~W.~Heinz, Phys.\ Rev.\  C {\bf 77}, 064901 (2008)
  [arXiv:0712.3715 [nucl-th]].


\bibitem{Song:2007fn}
H.~Song and U.~W.~Heinz,
  Phys.\ Lett.\  B {\bf 658}, 279 (2008)
  [arXiv:0709.0742 [nucl-th]].



\bibitem{chau}
A. Chaudhuri, J. Phys. \textbf{G35}, 104015 (2008).

\bibitem{du}
K. Dusling and D. Teaney, Phys. Rev. {\bf C 77}, 034905 (2008).

\bibitem{pasi} D. Molnar and P. Huovinen, J. Phys. \textbf{G35}, 104125
(2008).




\bibitem{pratt}
S. Pratt, Phys. Rev. \textbf{C77}, 024910 (2008).

\bibitem{bha}
R. S. Bhalerao and
S. Gupta, Phys. Rev. \textbf{C77} 014902 (2008).

\bibitem{mol}
I. Bouras, E. Molnar, H. Niemi, Z. Xu, A. El, O. Fochler, C. Greiner and D.H. Rischke,
Phys.\ Rev.\ Lett.\  {\bf 103}, 032301 (2009)
  [arXiv:0902.1927 [hep-ph]].


\bibitem{Betz:2008me}
  B.~Betz, D.~Henkel and D.~H.~Rischke,
 Prog.\ Part.\ Nucl.\ Phys.\  {\bf 62}, 556 (2009)
  [arXiv:0812.1440 [nucl-th]]; J.\ Phys.\ G {\bf 36}, 064029 (2009).



\bibitem{gue}
M. Martinez and M. Strickland, Phys. Rev. {\bf C 79}, 044903 (2009).

\bibitem{dkkm1}
T. Koide, G. S. Denicol, Ph. Mota and T. Kodama, Phys. Rev.
\textbf{C75}, 034909 (2007).

\bibitem{dkkm2}
G.S.Denicol, T. Kodama, T. Koide and Ph. Mota, Phys. Rev. {\bf C78},
034901 (2008).


\bibitem{dkkm3} G.S. Denicol, T. Kodama, T. Koide and Ph. Mota,
J. Phys. {\bf G 35}, 115102 (2008).



\bibitem{dkkm4}
G. S. Denicol, T. Kodama, T. Koide and Ph. Mota, J. Phys.
\textbf{G36}, 035103 (2009).


\bibitem{dkkm5}
G. S. Denicol, T. Kodama, T. Koide and Ph. Mota,
  Phys.\ Rev.\  C {\bf 80}, 064901 (2009)
  [arXiv:0903.3595 [hep-ph]].


\bibitem{knk}
T. Koide, E. Nakano and T. Kodama, Phys. Rev. Lett. {\bf 103}, 052301 (2009).







\bibitem{samojeden}
L.L. Samojeden and G.M. Kremer, Physica A {\bf 307}, 354 (2002).

\bibitem{is}
W.~Israel and J.~M.~Stewart, Ann. Phys. (N.Y.) \textbf{118},
341 (1979).



\bibitem{jou}
D.~Jou, J.~Casas-V\'azquez, and G.~Lebon,
Rep.~Prog.~Phys.~\textbf{51}, 1105 (1988);
\textbf{62}, 1035 (1999).

\bibitem{else}
M. Grmela and H. C. \"Ottinger, Phys. Rev. {\bf E 56},
6620 (1997); B.
Carter, Proc. R. Soc. {\bf A433}, 45 (1991).


\bibitem{baier} R.~Baier, P.~Romatschke, D.~T.~Son,
A.~O.~Starinets and M.~A.~Stephanov, JHEP {\bf 0804}, 100 (2008).


\bibitem{Xu}
  Z.~Xu and C.~Greiner,
  Phys.\ Rev.\ Lett.\  {\bf 100}, 172301 (2008)
  [arXiv:0710.5719 [nucl-th]].





\bibitem{his}W. A Hiscock and L. Lindblom, Ann. Phys. (N.Y.) {\bf 151},
466 (1983); Phys. Rev. {\bf D31}, 725 (1985);
{\it ibid} {\bf D35}, 3723 (1987);
Phys. Lett. {\bf A131}, 509 (1988);
W. A. Hiscock and T. S. Olson, Phys. Lett. {\bf A141}, 125 (1989).

\bibitem{his2}
W. A Hiscock and L. Lindblom, Phys. Rev. {\bf D31}, 725 (1985);
{\it ibid} {\bf D35}, 3723 (1987).



\bibitem{Heller:2007qt}
M.~P.~Heller and R.~A.~Janik, Phys.\ Rev.\  D {\bf 76}, 025027 (2007).


\bibitem{push}
S.~Pu and Q.~Wang, arXiv:0810.5271 [hep-ph].

\bibitem{jackson}
J. D. Jackson: {\it Classical Electrodynamics},
(John Wiley $\&$ Sons, Inc., 1999).
\bibitem{bri}
L. Brillouin: {\it Wave Propagation and Group Velocity},
(Academic Press, London, 1960).






\end{thebibliography}
\end{document}